\documentclass[twocolumngrid,prd,superscriptsize,reprint]{revtex4-1}
\usepackage[utf8]{inputenc} 
\usepackage{graphicx,xcolor,overpic,mathtools}
\usepackage{amsthm,amsmath,amssymb}
\usepackage[colorlinks]{hyperref}
\usepackage{textcomp}
\definecolor{coolblack}{rgb}{0.0, 0.18, 0.39}
\hypersetup{
    colorlinks = false,
   }
\PassOptionsToPackage{normalem}{ulem}
\usepackage{ulem} 
\usepackage{sidenotes}
\usepackage{bm}
\usepackage{url}
\usepackage{physics}
\usepackage[makeroom]{cancel}
\usepackage{cleveref}
\usepackage{tensor}
\usepackage{mathrsfs}
\usepackage{slashed}
\usepackage{caption}
\usepackage{subcaption}
\usepackage[mode=buildnew]{standalone}

\newcommand{\comment}[1]{}

\usepackage{pgfplots}

\usepackage{diagbox}
\usepackage{stackengine,amssymb,graphicx}

\NewDocumentCommand{\evat}{sO{\bigg}mm}{%
  \IfBooleanTF{#1}
   {\mleft. #3 \mright|_{#4}}
   {#3#2|_{#4}}%
}
\usepackage{enumerate}
\definecolor{azure}{rgb}{0.0, 0.5, 1.0}

\begin{document}

\title[]{Universal relations for rotating Boson Stars}

\author{Christoph Adam}
\author{Jorge Castelo}
\author{Alberto Garc\'ia Mart\'in-Caro}
\author{Miguel Huidobro}%
\author{Ricardo V\'azquez}
\affiliation{%
Departamento de F\'isica de Part\'iculas, Universidad de Santiago de Compostela and Instituto
Galego de F\'isica de Altas Enerxias (IGFAE) E-15782 Santiago de Compostela, Spain
}%
\author{Andrzej Wereszczynski}
\affiliation{
Institute of Physics, Jagiellonian University, Lojasiewicza 11, Krak\'ow, Poland
}%

\date[ Date: ]{\today}
\begin{abstract}

Boson stars represent a hypothetical exotic type of compact stellar object that may be observed from the gravitational signal of coalescing binaries in current and future GW detectors.
In this work we show that the moment of inertia $I$, the (dimensionless) angular momentum $\chi$ and the quadrupole moment $Q$ of rotating boson stars obey a universal relation, valid for a wide set of boson star models. Further, the obtained $I-\chi-Q~$ relation clearly differs from its famous neutron star counterpart, providing us with an unequivocal diagnostic tool to distinguish boson stars from ordinary compact stars or other celestial bodies in GW observations. 
Such universal (i.e. model-independent) relations also provide a useful tool to probe the strong gravity regime of general relativity and to constrain the equation of state of matter inside compact stars.

\end{abstract}
\maketitle
\tableofcontents

\section{Introduction.}

Boson stars (BSs) are localised solutions of a bosonic field theory (in the simplest case just a complex scalar $\Phi$) coupled to 
gravity (for reviews see \cite{Liebling:2012fv,Lai:2004fw,Schunck:2003kk}). After the first \textit{geons} proposed by Wheeler \cite{PhysRev.97.511}, the original spherical scalar BS was introduced by Kaup \cite{PhysRev.172.1331}, Ruffini, Bonazzola and Pacini \cite{PhysRev.187.1767,PhysRev.148.1269}. It is now well understood that 
properties of BSs strongly depend on the form of the Lagrangian, i.e., the potential which encodes the self-interaction of 
the complex scalar. Various kinds of potentials allowed to model a large range of astrophysical objects. E.g., there are BSs with 
properties very similar to neutron stars (NSs) or to black holes (BHs) (black-hole-mimickers \cite{PhysRevD.80.084023}). Some BSs
are even candidates for dark-matter galaxy halos \cite{Schunck:1998nq}. 

In the last forty years, important BS generalizations have been found like, e.g., the vector scalar field solutions, called Proca stars 
\cite{Brito:2015pxa}. They have also been extended to generalized models of gravity like Einstein-Gauss-Bonnet theory, scalar-tensor models \cite{PhysRevD.56.3478} or Palatini gravity \cite{Maso-Ferrando:2021ngp}. 
Further, the stability of such self gravitating scalar field solutions has  been investigated, e.g., in \cite{Khlopov:1985jw, Sanchis-Gual:2019ljs, Sanchis-Gual:2021phr,Siemonsen:2020hcg,DiGiovanni:2020ror}. 

The interest in this topic has increased considerably in the last decade. On the one hand, the discovery of the fundamental scalar Higgs boson at CERN \cite{ATLAS:2012yve,CMS:2012qbp} 
provides a convincing argument for the possible existence of further scalar fields extending the Standard Model, such as the axion \cite{PhysRevLett.40.223,PhysRevLett.40.279} or other ultralight scalar or vector bosons \cite{Freitas:2021cfi} postulated as potential dark-matter particles.  
  On the other hand, gravitational wave (GW) astronomy provides a new tool for the search of exotic compact objects such as BSs. Since LIGO and Virgo reported the first event \cite{LIGOScientific:2016aoc}, where the GW
signal from a BH binary merger was measured, the current observatories, such as advanced LIGO, advanced Virgo
or KAGRA have reported more than forty events \cite{LIGOScientific:2021hvc,KAGRA:2018plz}, among which binary NS \cite{LIGOScientific:2018cki}, binary BH and even NS-BH mergers have been identified. In this promising scenario, a very particular GW signal was measured in 2020 by advanced LIGO-Virgo which could be potentially explained as a head-on collision of two Proca stars \cite{Bustillo:2020syj}. Binary BS mergers are also being studied nowadays \cite{Bezares:2022obu}.

Unlike regular, perfect fluid stars, rotating BSs differ a lot from their static counterparts. For example, it is not possible to obtain slowly rotating BSs as a perturbation of the static solution \cite{Kobayashi:1994qi} within the standard Hartle-Thorne formalism \cite{Hartle:1967he,Hartle:1968si}. Still, rotating BS solutions do exist as proved by Silveira \& de Sousa \cite{Silveira:1995dh}, following work of  Ferrell \& Gleiser  \cite{Ferrell:1989kz}, but they require a nonperturbative treatment, and cannot be understood as a rigidly rotating system. This makes analytical and numerical computations more involved, if compared with other relativistic compact objects such as NSs or BHs.

For rotating NSs, a very important and not yet fully explained property is the existence of universal relations which do not depend on their equation of state (EOS). The most famous set of such relations are the so-called $I$-
Love-$Q$ relations, proposed by Yagi and Yunes in \cite{Yagi:2013awa}, involving the moment of inertia $I$, the tidal deformability (Love number) \cite{Hinderer:2007mb,Postnikov:2010yn} 
and the quadrupolar moment $Q$.

It is quite challenging to test these relations and extract $I, Q$ and the spin from astrophysical data, but there exist some promising possibilities.  The double
pulsar J0737-3039 is expected to provide the first direct NS measurement of the moment of inertia \cite{Kramer:2009zza}, and some estimations have already been given using NICER measurements \cite{Silva:2020acr}. Other methods for measuring the moment of inertia rely on GW observations \cite{Silva:2016myw}, glitches measurements \cite{Link:1999ca,Andersson2012PulsarGT,Chamel2013CrustalEA}, etc. \cite{Steiner:2014pda,Damour:1988mr,Lattimer:2004nj,Bejger:2005jy}. The future oscillation mode measurements by the GW community should allow to obtain the quadrupolar moments $Q$ \cite{Zhao:2022tcw}, and there exist already some current estimations for $Q$  \cite{abbott2017gw170817,Silva:2020acr}.

Since their discovery, these relations have been extended to more realistic situations including a high spin velocity and magnetic fields \cite{Haskell:2013vha}, and to modified gravity theories
\cite{Sham:2013cya,Chakravarti:2019aup,Doneva:2017jop}. It is now well established that they hold for realistic EOS in the slow rotation limit \cite{Adam:2020aza}. Following these results, other quasi-universal relations, involving higher multipoles and Love numbers \cite{Yagi:2013sva,Godzieba:2021vnz}, the compactness, gravitational binding energies \cite{UnivRelsbinding} , and oscillation frequencies of (quasi) normal modes \cite{Torres-Forne:2019zwz}
both for slowly and fast spinning NSs have been studied \cite{Sun:2020qkj,Doneva:2017jop}.

These universal relations are very important for several reasons. If the nature of a star is known, then multiple observations would allow us to verify the validity of these relations or, assuming their validity, to determine further properties of that star which cannot be observed directly.

If the nature of a compact star is not known, on the other hand, then universal relations which clearly distinguish between different types of stars -- like, e.g., NSs and BSs -- would in principle allow to determine its nature or to eliminate certain possibilities. This requires, however,  that multiple observations allow for an independent determination of different observables (different multipole moments) with sufficient precision.

The aim of the current work is, therefore, to analyze the existence of universal relations for rotating BSs. We emphasize that, although certain BS models can mimick NSs or BHs, they correspond to quite different  types of solutions. It is, therefore, an important open question whether such universal relations exist for BSs and, if they exist, whether they are identical to the relations found for rotating NSs or whether these two types of compact objects obey rather distinct universal laws.


We will use in what follows $\hbar=c=1$.


\section{Theoretical set-up.}

We start with the Einstein-Klein-Gordon (EKG) action describing a massive complex scalar field $\Phi$ minimally coupled to Einstein gravity \cite{Liebling:2012fv},
\begin{equation}
    \mathcal{
    S}=\int \left(\frac{1}{16\pi G}R+\mathcal{L}_{\Phi}\right)\sqrt{-g}d^4x.
    \label{action}
\end{equation}
Here $g$ is the metric determinant, and $R$ the Ricci scalar. The Lagrangian governing the complex field dynamics reads,
\begin{equation}
 \mathcal{L}_{\Phi}=-\frac{1}{2}\left[g^{\alpha\beta}\nabla_{\alpha}\Phi^*\nabla_{\beta}\Phi+V\left(|\Phi|^2\right)\right],
    \label{lagrangian}
\end{equation}
where $V\left(|\Phi|^2\right)$ is a potential that depends only on the absolute value of the scalar field, respecting the global $U(1)$ invariance of the model. All potentials we consider contain the quadratic mass term $\mu^2|\Phi|^2$, whereas higher self-interaction terms will vary significantly. 

Varying the action (\ref{action}) yields the EKG equations,
\begin{equation}
\begin{split}
    &R_{\alpha\beta}-\frac{1}{2}Rg_{\alpha\beta}=8\pi T_{\alpha\beta}, \\
    &
    g^{\alpha\beta}\nabla_\alpha\nabla_{\beta}\Phi=\frac{dV}{d|\Phi|^2}\Phi,
\label{kg}
\end{split}
\end{equation}
where $R_{\alpha\beta}$ is the Ricci tensor and $T_{\alpha\beta}$ is the canonical Stress-Energy tensor of the scalar field,
\begin{equation}
    T_{\alpha\beta}=2\nabla_{(\alpha}\Phi^*\nabla_{\beta)}\Phi-2g_{\alpha\beta}\left[g^{\mu\nu}\nabla_{(\mu}\Phi^*\nabla_{\nu)}\Phi+V\left(|\Phi|^2\right)\right]].
    \label{stress}
\end{equation}
Rotating compact objects lead in a natural way to axially symmetric systems. Therefore, we assume the following stationary, axially symmetric ansatz for the metric \cite{Herdeiro:2015gia,PhysRevD.55.6081},
\begin{equation}
\begin{split}
    ds^2=&-e^{2\nu}dt^2+e^{2\beta}r^2\sin^2\theta\left(d\psi-\frac{W}{r}dt\right)^2\\
    &+e^{2\alpha}(dr^2+r^2d\theta^2),
    \label{Herdeiro}
    \end{split}
\end{equation}
where $\nu, \alpha, \beta$ and $W$ are functions dependent only on $r,\theta$. Furthermore, the consistent ansatz for the scalar field is,
\begin{equation}
     \Phi(t,r,\theta,\psi)=\phi(r,\theta)e^{-i(w t+n\psi)}.
     \label{scalar}
 \end{equation}
 Here, $\phi(r,\theta)$ is the modulus of the complex field, usually referred to as the profile of the star.
 Further, 
 $w \in \mathbb{R}$ is the angular frequency of the field and $n \in \mathbb{Z}$ is the azimutal harmonic index.
 All calculations in this work are done for the fixed value $n=1$, because this is the most paradigmatic, most studied and simplest case. Numerical calculations for higher $n$ need the solutions for lower $n$ as initial data. Nevertheless, we have performed some preliminary higher $n$ calculations, and the results are that {\em i)} for low $n>1$, each $n$ defines its own universal surface in the 
$I,\chi ,Q$ space, {\em ii)} a limiting universal surface is approached for large $n$, and {\em iii)} all these universal surfaces are clearly different from the NS case. A full presentation and discussion of more detailed higher $n$ calculations, as well as several further results, will be provided in a forthcoming publication.

\subsection{Boson star models.}


The properties of different BSs, i.e., solutions of the EKG system,
strongly depend on the potential. The scalar potential for BS solutions plays an analogous role to the EOS in the case of NSs.

\begin{figure}[h!]
\centering
\includegraphics[width=0.5\textwidth]{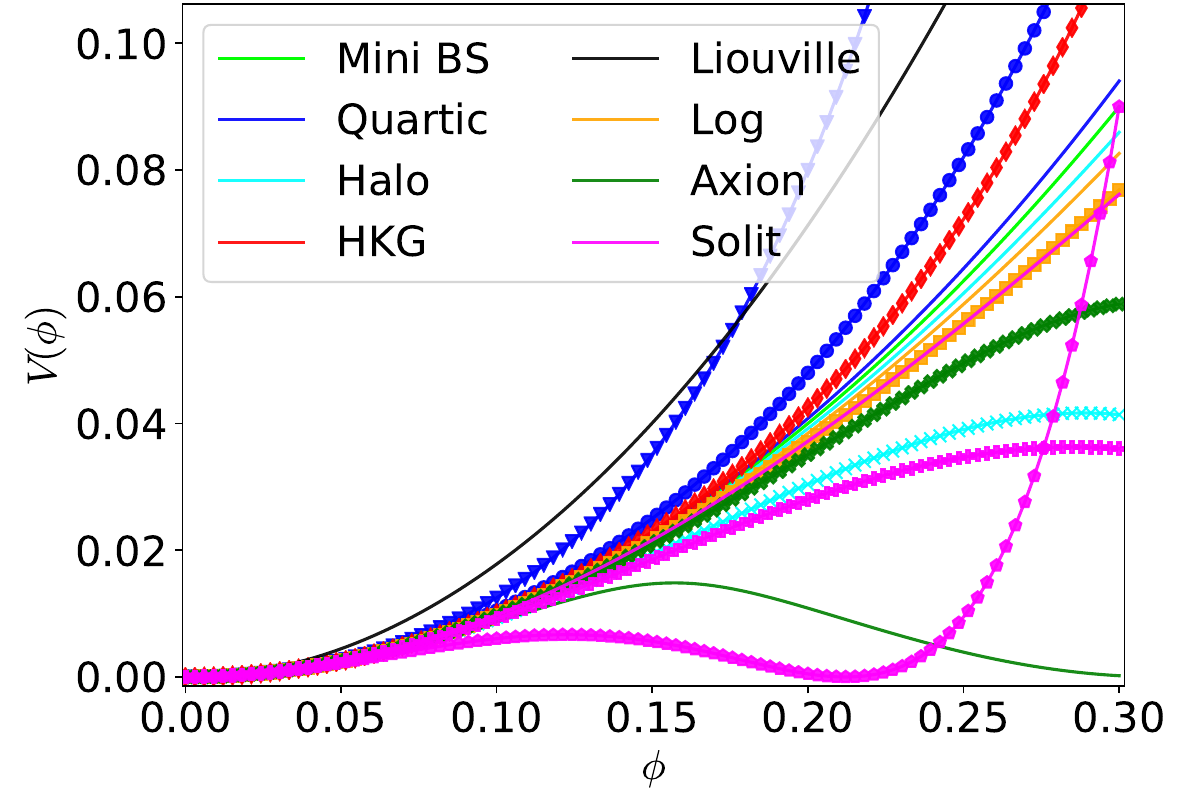}
\caption{
Form of the potentials $V(\phi)$, in the relevant field range $\phi\in [0,0.29]$ (in the numerical simulations the field always takes values within this range). We can appreciate rather different behaviors, resulting in rather different BSs. Each color denotes a different model, and different symbols correspond to different parameter values within a model. These parameter values are given in the supplementary material. }
\label{fig:x cubed graph}
\end{figure}

As we are interested in obtaining universal properties of BS solutions, in this paper we have selected a set of potentials which {\it i)} is physically well-motivated, i.e., the resulting ranges of masses, radii and compactness fit to various astrophysical scenarios, like dark matter haloes, \cite{Mielke:2019rvl,Mielke:2020mve} or to BH and NS-like objects \cite{Choi:2019mva,Guerra:2019srj,Delgado:2020udb,Vaglio:2022flq,Grandclement:2014msa}, and {\it ii)} covers a wide range of potentials considered in the literature with rather different qualitative features, see \cref{Table.Potentials} and \cref{fig:x cubed graph} for details.
Some of these models present instabilities (i.e the so-called Mini Boson Star \cite{Khlopov:1985jw, Sanchis-Gual:2021phr,Siemonsen:2020hcg,DiGiovanni:2020ror}), but we also treat them here because we want to find the most general behavior in this work.

The simplest choice is just a mass term without any self-interaction, the so-called \textit{Mini-boson star} potential. This can be further generalized with the inclusion of higher order self-interaction terms, e.g. $|\Phi|^4$ and $|\Phi|^6$ \cite{Schunck:2003kk,Colpi:1986ye,Grandclement:2014msa}. Finally, potentials based on the logarithm, exponential and sine functions (Axion potential) have also been considered \cite{Delgado:2020udb,Choi:2019mva,Guerra:2019srj}.

\begin{table}[h!]
	\centering
		\begin{tabular}{|c|c|}
			\hline
		 Name & $V\left(\phi\right)$ \\ \hline
			Mini-BS, BS$_{\rm Mass}$& $V_{\rm Mass}=\mu^2\phi^2$ \\ \hline
		BS$_{\rm Quartic}$&$V_{\rm Quartic}=\mu^2\phi^2+\lambda/2\phi^4$    \\  \hline
				BS$_{\rm Halo}$& $V_{\rm Halo}=\mu^2\phi^2-\alpha\phi^4$  \\ \hline
			BS$_{\rm HKG}$ & $V_{\rm HKG}=\mu^2\phi^2-\alpha\phi^4+\beta\phi^6$  \\ \hline
			BS$_{\rm Sol}$& $V_{\rm Sol}=\mu^2\phi^2(1-(\phi^2/\phi_0^2))^2$  \\ \hline
			BS$_{\rm Log}$&$V_{\rm Log}=f^2\mu^2 \ln\left(\phi^2/f^2+1\right)$\\ \hline
			BS$_{\rm Liouville}$& $V_{\rm Liouville}=f^2\mu^2 \left(\exp\{\phi^2/f^2\}-1\right)$  \\ \hline
				BS$_{\rm Axion}$& $V_{\rm Axion}=\frac{2\mu^2f^2}{B}\left(1-\sqrt{1-4B\sin^2(\phi/2f)}\right)$  \\
			\hline
		\end{tabular}
		\caption{\small BS potentials analyzed in the current work.}
		\label{Table.Potentials}
\end{table}

All of these potentials have been previously considered in the case
of spherical, non-rotating BSs, which was, in some cases, further
generalized to rotating solutions   \cite{Siemonsen:2020hcg}.

\section{Numerical implementation. }


To perform the numerical integration of the EKG system we first rescale the radial distance and angular frequency by the mass $\mu$ of the boson field, $
r\rightarrow r\mu, \hspace{0.2cm}w\rightarrow w/\mu$, thus removing the explicit $\mu$ dependence from the field equations. 
For simplicity, we also rescale the field $ \phi\rightarrow\phi\sqrt{4\pi}$. 

The mathematical problem we have to solve is a set of five coupled, non-linear, partial differential equations for the metric functions and the scalar field, which follows from (\ref{kg}). We also take into account the constraints, $E^r_{\theta}=0, E^r_r-E^{\theta}_{\theta}=0$, where $E^{\mu}_{\nu}=R^{\mu}_{\nu}-\frac{1}{2}Rg^{\mu}_{\nu}-2 T^{\mu}_{\nu}$.

To perform the numerical integration, we used the FIDISOL/CADSOL solver \cite{fidisol,schonauer1989efficient,schonauer2001we}, a professional package developed at the Karlsruhe Institute of Technology in the eighties. The package is built in Fortran 90 and solves nonlinear, two or three-dimensional, elliptic and parabolic partial differential equations (PDEs). It is a Newton-Raphson based finite difference code. It allows us to use arbitrary boundary conditions and works on a rectangular domain, with a self-adaptative grid and consistency order.
As a method based on finite differences, it finds the function's roots.
Following \cite{Delgado:2022pwo}, let us explain the basics:
\begin{enumerate}
    \item The system of equations has to be written as follows,
    \begin{equation}
        P(x,y;u,u_{x},u_{y},u_{xx},u_{yy},u_{xy})=0,
    \end{equation}
    where $x,y$ are the independent variables, $u$ the set of functions to be solved and $\{u_{x},u_{y},u_{xx},u_{yy},u_{xy}\}$ the derivatives of $u$ with respect to the given sub-indices.
    \item The package needs also the Jacobi equations for all the functions, i.e, the derivatives with respect to $\{u,u_{x},u_{y},u_{xx},u_{yy},u_{xy}\}$.
    \item We need to provide the solver with an accurate initial guess and boundary conditions.
    \item Finally, we have to choose the grid over  which the equations are discretized. The mesh for the variables $x,y$ (here $x=r/(1+r) \in [0,1]$ and $y = \theta \in [0,\pi/2]$) has a number of points $N_x,N_y$. We could make variations of this number of points, and within a reasonable margin of variation, we still should reach the correct solution. This means that small changes on the grid are not translated into instabilities. For most of our calculations we used a $401\times 40$ grid, but we will comment on the effect of other grids for the particular case of the mini-boson star potential, see below.
    
\end{enumerate}
Let us now explain the basics of the numerical approach of the solver.
\begin{itemize}
    \item To start, we have to provide the solver with an initial guess $u^{(1)}$ for which $P(u^{(1)})\neq 0$. This difference cannot be too large, as we want the program to converge. It is here where it is important  have a good initial guess.
    \item The new, improved guess is introduced in the following way,
    \begin{equation}
        u^{(2)}=u^{(1)}+s\Delta u^{(1)},
        \label{improved}
    \end{equation}
    where $s$ is a relaxation parameter, usually set equal to $s=1$. 
    \item The next step is to expand $P(u^{(2)})$ up to first order in the small parameter $\Delta u^{(1)}$ and to assume that $P(u^{(2)})=0$ to this order,
    \begin{eqnarray}
     P(u^{(2)})&=&P(u^{(1)}+ \Delta u^{(1)}) \\ \nonumber
     &\cong&P(u^{(1)})+\frac{\partial P}{\partial u}(u^{(1)})\Delta u^{(1)}\cong 0.
    \end{eqnarray}
\item     Now we compute $\Delta u^{(1)}$ through the above equation and obtain a new improved approximation $u^{(2)}$ by using \cref{improved}.
 As $P(u^{(1)})$ and $\Delta u^{(1)}$ are vectors and $\frac{\partial P}{\partial u}(u^{(1)})$ is a matrix, the program solves a typical linear system $Ma=b$.
    \item Then the process is repeated iteratively to get
    $$ u^{(3)}=u^{(2)}+\Delta u^{(2)}, u^{(4)}=u^{(3)}+\Delta u^{(3)}$$ and so on, until the Newton Residual $P(u^{(N)})$ reaches a value lower than the desired tolerance.
\end{itemize}
After many iterations, the change of the Newton residuals in consecutive iterations becomes lower than the tolerance, meaning that the system is solved. The number of iterations slightly varies, depending on the model and the frequency, but is always of the order of $10^4$. Further, the package provides an error estimation for each function, computed through the discretized Newton residual and the errors of the function's derivatives. The discretization is done by using the backward finite difference method \cite{10.2307/88864}, up to an arbitrary consistency order, chosen by the user. We choose the tolerance (an internal parameter of the code) in such a way that the resulting errors for the metric potentials and the scalar field are always lower than $10^{-3}$.

The solver needs the equations in a specific form, so we need to work with certain combinations of the Einstein equations $ E^{\nu}_{\mu}\equiv G^{\mu}_{\nu}-2\kappa^2T_{\mu}^{\nu}=0$  and the Klein-Gordon equation, to find a set of $5$ independent equations, such that each equation only contains {\em one} field with second order derivatives (here "field" refers to the metric potentials and the scalar field). We also multiply the equations by suitable factors to get rid of numerical divergences like $1/r$ or $1/\sin(\theta)$, resulting in the generic form ($i=1 ,\ldots 5$)
\begin{equation}
\begin{split}
   &r^2\sin^2{\theta}F_{i,rr} + \, \sin^2{\theta}F_{i,\theta\theta}\; + \\
   &\mathcal{F}_i(r,\theta;F_j(r,\theta);\partial {F}_j(r,\theta))\, = \, 0.  
\end{split}
\end{equation}
Here $F_i(r,\theta)$ are the different fields $\{\nu,\alpha,\beta,W,\phi\}$, and $\mathcal{F}_i$ is the sum of the remaining terms, containing only the fields and their first derivatives with respect to $r$ and $\theta$. 
So our EKG system is presented in the following form:
\begin{equation}
\begin{split}
&-e^{2\alpha}\frac{r^2}{2}\sin^2(\theta)\left(-E_{t}^{t}+E_{r}^{r}+E_{\theta}^{\theta}-E_{\phi}^{\phi}\right)=0\\
&e^{2\alpha}\frac{r^2}{2}\sin^2(\theta)\left(E_{t}^{t}+E_{r}^{r}+E_{\theta}^{\theta}-E_{\phi}^{\phi}+\frac{2WE_{\phi}^{t}}{r}\right)=0\\
&e^{2\alpha}\frac{r^2}{2}\sin^2(\theta)\left(-E_{t}^{t}+E_{r}^{r}+E_{\theta}^{\theta}-E_{\phi}^{\phi}-\frac{2WE_{\phi}^{t}}{r}\right)=0\\
&2re^{2\nu +2\alpha-2\beta}E_{\phi}^{t}=0\\
& \frac{e^{2\alpha}r^2\sin^2(\theta)}{\phi} \Phi^*\left(\Box-\frac{d V}{d|\phi|^2} \right) \Phi=0.
\end{split}
\label{EKG-system}
\end{equation}



 Finally, we have to impose boundary conditions on the field profile and the metric functions. Asymptotic flatness implies that all of them must vanish at infinity.
Axial symmetry together with reflection on the rotation axis implies that at $\theta=0$ and $\theta=\pi$, the metric functions and the profile also go to $0$.

Symmetry with respect to a reflection along the equatorial plane implies that these functions also have to vanish at $\theta=\pi/2$.
Finally, regularity at the origin requires that $\partial_r \alpha=\partial_r \beta=\partial_r \nu =W= \phi=0$ for $r \to 0$, and regularity in the symmetry axis further imposes $\left.  \alpha=\beta \right|_{\theta=0, \pi}$ \cite{Herdeiro:2015gia}.


It is important to remember that the field profiles of static and rotating BSs are qualitatively very different. Indeed, they have different topologies. While static BSs are spherically symmetric with the central value of the complex field being a free parameter of the solution, rotating BSs have a toroidal shape with vanishing scalar field on the rotation axis, $\theta =0$.
  Near the axis, it behaves like
\begin{equation}
    \lim_{r\rightarrow 0}\phi(r,\theta)=r^{n}h_n(\theta)+O(r^{r+2}),
\end{equation}
where $h_n$ are particular functions, different for each harmonic index $n$. 


\section{Multipolar structure and global properties.}


In this section we derive the principal quantities which constitute the universal relation, i.e., the moment of inertia $I$ and the quadrupole moment $Q$. This will require the multipole expansion. Strictly speaking, however, BSs are infinitely extended objects without any particular surface \cite{Liebling:2012fv}. Simply, the scalar field extends to arbitrarily large distances. Following \cite{Delgado:2020udb}, we identify radii with the perimetral radius that contains $99\%$ of the BS matter. 

\subsection{Multipole moments}

In the derivation of multipole moments we follow the procedure developed in \cite{Pappas:2018csu,1976ApJ...204..200B} for NSs. We introduce a new parametrization of the metric \cref{Herdeiro}, defining the new metric functions $\omega=\frac{W}{r},\hspace{0.4cm}B=e^{\nu+\beta}$.
The following expressions provide a consistent asymptotic multipolar expansion of the metric functions (see \cite{1976ApJ...204..200B,morse1954methods}),
\begin{eqnarray}
        \nu&=&\sum_{l=0}^{\infty}\bar{\nu}_{2l}(r)P_{2l}(\cos\theta),\hspace{0.4cm}   \hspace{0.9cm} \bar{\nu}_{2l}(r)= \sum_{k=0}^{\infty}\frac{\nu_{2l,k}}{r^{2l+1+k}}, \nonumber  \\
         \omega&=&\sum_{l=0}^{\infty}\bar{\omega}_{2l-1}(r)\frac{dP_{2l-1}(\cos\theta)}{d\cos\theta},     \hspace{0.2cm} \bar{\omega}_{2l-1}(r)= \sum_{k=0}^{\infty}\frac{\omega_{2l-1,k}}{r^{2l+1+k}} \nonumber  \\
          B&=&1+\sum_{l=0}^{\infty}\bar{B}_{2l}(r)T_{2l}^{\frac{1}{2}}(\cos\theta),     \hspace{0.2cm}     \hspace{0.2cm} \bar{B}_{2l}(r)= \frac{B_{2l}}{r^{2l+2}},       
    \label{multipoles}
\end{eqnarray}
where $P_l(\cos\theta)$ and $T_l^{\frac{1}{2}}(\cos\theta)$ are the Legendre and Gegenbauer polynomials, respectively. Then, multipole moments can be found as combinations of the expansion coefficients in (\ref{multipoles}), see \cite{Pappas:2014gca} for details. Specifically, one can show that the mass monopole $M$, angular momentum dipole $J$ and quadrupole $Q$ moments  are, 
\begin{equation}
\begin{split}
 &M = -\nu_{0,0},\quad J =\frac{\omega_{1,0}}{2},\\
 &Q=\frac{4}{3}B_{0}\nu_{0,0}+\frac{\nu_{0,0}^3}{3}-\nu_{2,0},
    \label{M1}
    \end{split}
\end{equation}

As these coefficients are crucial in our analysis, let us explain how we obtain them. Instead of the full source integration (see, e.g.,  \cite{PhysRevD.55.6081}, \cite{Doneva:2017jop}), we use the fact that we already solved the EKG system numerically and, hence, know the functions $\nu$, $\omega$ and $B$. The multipole coefficients are then found by integrating over the angles after projecting on the appropriate polynomial, and taking the pertinent radial limits.
We explicitly find,
\begin{equation}
\begin{split}
&
\nu_{0,0}=\frac{1}{2}\lim_{r\rightarrow\infty}r\int_{-1}^{1}\nu(r,\theta)d\cos\theta,\\
&
\nu_{2,0}=\frac{5}{2}\lim_{r\rightarrow\infty}r^{3}\int_{-1}^{1}\nu(r,\theta)\frac{(3\cos^2\theta-1)}{2}d\cos\theta,\\
&
\omega_{1,0}=\frac{1}{2}\lim_{r\rightarrow\infty}r^{3}\int_{-1}^{1}\omega(r,\theta)d\cos\theta,\\
&
B_{0}=\lim_{r\rightarrow\infty}r^{2}\int_{-1}^{1}(B(r,\theta)-1)\sin\theta\sqrt{\frac{2}{\pi}}d\cos\theta.
    \label{coeffint}
\end{split}
\end{equation}

We focused on the observables related to the lowest multipole moments, because the chances that they could be determined by observations in the not-too-distant future are higher. But in principle higher moments (like the octupole moment) can be calculated without difficulties by our methods, and the possible existence of further universal relations can be investigated, similarly to what was done, e.g., in \cite{Pappas:2013naa} for the NS case.

We also compared the mass and angular momentum obtained from \cref{M1} with those obtained from the Komar integrals \cite{PhysRev.129.1873}, and we found a good agreement with less than $2\%$ discre\-pancy.

\subsection{Moments of inertia and differential rotation}

Rotating NSs are often assumed to be rigidly rotating objects, whose moment of inertia $I$ is defined as, 
\begin{equation}
    I=\frac{J}{\Omega},\hspace{0.2cm}\text{where}\hspace{0.2cm}\Omega=\frac{d\psi}{dt}=\frac{u^{\psi}}{u^t},
    \label{NSangular}
\end{equation}
being $u^\mu$ the four-velocity of an observer comoving with the fluid. For spinning BSs, instead, it is not obvious how to obtain the four-velocity, as the corresponding stress energy tensor (\cref{stress}) for a complex scalar cannot be rewritten in a perfect fluid form. This differs from the real scalar field case, where the stress-energy tensor can indeed be brought to this form \cite{Faraoni:2012hn}.
Moreover, under the strong-coupling assumption proposed by Ryan in \cite{PhysRevD.55.6081}, in which $\partial_r\phi $ and $\partial_\theta \phi$ can be neglected, the tensor \eqref{stress} acquires a perfect fluid form with a barotropic equation of state \cite{PhysRevD.55.6081}. This approximation was recently used in \cite{Vaglio:2022flq} to study the multipolar structure of rotating BSs. 

We shall use another strategy to find a well-defined moment of inertia that does not rely on any approximation, taking advantage of the fact that there is a natural four-vector associated with the global $U(1)$ symmetry of the Lagrangian, i.e., the corresponding Noether current,
\begin{equation}
    j^{\mu}=\frac{i}{2}\sqrt{|g|}g^{\mu\nu}\left[\Phi^*\nabla_{\nu}\Phi-\Phi\nabla_{\nu}\Phi^*\right],
\end{equation}
which gives rise to the conserved particle number $N=\int  j^0 \sqrt{-g}d^3x$.
Now, we define the differential angular velocity as, 
\begin{equation}
\begin{split}
    \Omega=\frac{j^{\psi}}{j^{t}}=\frac{wg^{\psi t}-ng^{\psi\psi}}{wg^{tt}-ng^{t\psi}}=\frac{W}{r}+\frac{ne^{2(\nu-\beta)}}{r^2\left(w-\frac{nW}{r}\right)\sin^2\theta}.
\label{diferentialfrequency}
\end{split}
\end{equation}
Remarkably, the expression in \cref{diferentialfrequency} agrees with that obtained by Ryan in \cite{PhysRevD.55.6081} in the strong coupling approximation. This proves that our definition, which is completely general, is consistent with Ryan's formula.

As a consequence of the differential rotation law \eqref{diferentialfrequency}, $\Omega$ is a function of $r$ and $\theta$. This must be taken into account when we compute the inertia tensor. Therefore, for a differentially rotating system we use the following expression which generalizes \cref{NSangular},
\begin{equation}
    I=\int_0^{\pi}\int_0^{\infty}\frac{j(r,\theta)}{\Omega(r,\theta)}r^2\sin\theta e^{\nu+2\alpha+\beta}drd\theta,
\end{equation}
where $j(r,\theta)=T^t_{\psi}$ is the angular momentum density.

\section{Universal relations.}
Once the multipoles have been obtained, we may define the standard dimensionless \emph{reduced multipole moments}
\cite{2013Sci...341..365Y},
\begin{equation}
    \bar{I}=\frac{I}{M_{99}^3},\hspace{0.2cm}\bar{Q}=\frac{Q}{M_{99}^3\chi^{2}},\hspace{0.1cm}\chi=\frac{J}{M_{99}^2},
    \label{reduced}
\end{equation}
where $M_{99}$ is $99\%$ of the total mass, and $\chi$ the dimensionless spin parameter. Naively trying to find $I-Q$ relations, as in the slowly rotating NS case, we find that these relations are not accurate, with a maximum difference of about 25\%.

\begin{figure}
\centering
\begin{subfigure}{0.4\textwidth}
    \includegraphics[width=\textwidth]{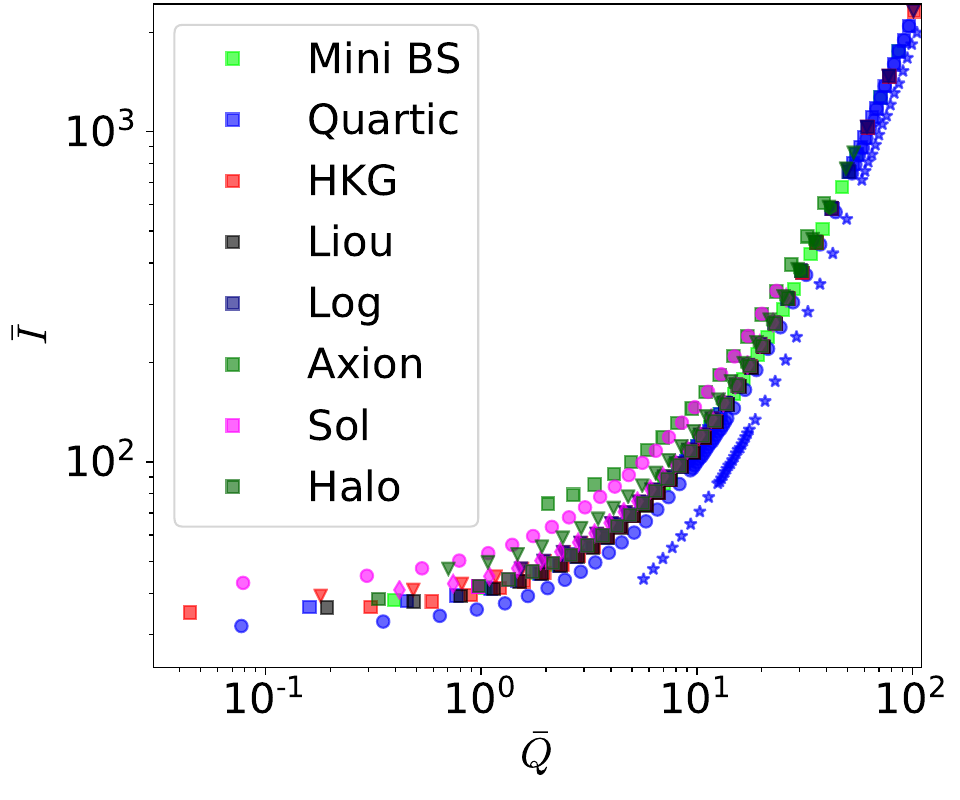}
    \label{fig:first}
\end{subfigure}
\begin{subfigure}{0.4\textwidth}
    \includegraphics[width=\textwidth]{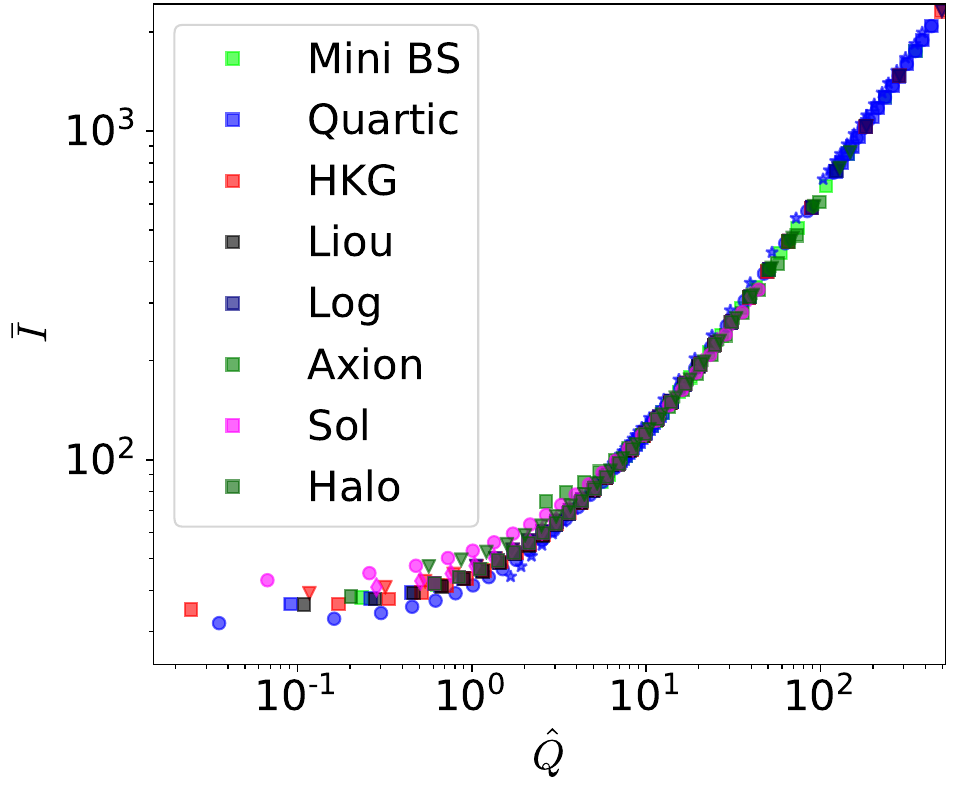}
    \label{fig:third}
\end{subfigure}
\caption{In the upper plot we show the moment of inertia vs. quadrupole moment, for the the usual reduced quadrupole moments $\bar{Q}$. In the lower plot we show the plot for the rescaled quadrupole moment $\Hat{Q}=Q\chi^{0.7}$ .}
\label{naiveplots}
\end{figure}

We can see in \cref{naiveplots} that the curves for different models diverge less if we use $\Hat{Q}=Q\chi^a$ where $a\in[0.51,0.75]$, instead of $\bar{Q}$. We tried with several $\chi$ power laws, and for $\chi^a$ with $a$ in the interval given above, the behavior improves with respect to the standard reduced dimensionless multipoles, but the differences for a global fitting are still too high. Concretely, we improve from a $25\%$ deviation in the usual multipoles to a $13\%$ using this new rescaling.

\begin{figure*}[]
\vspace*{-1.5cm}
\subfloat{%
  \hspace*{1.0cm}\includegraphics[clip,width=1.0\columnwidth]{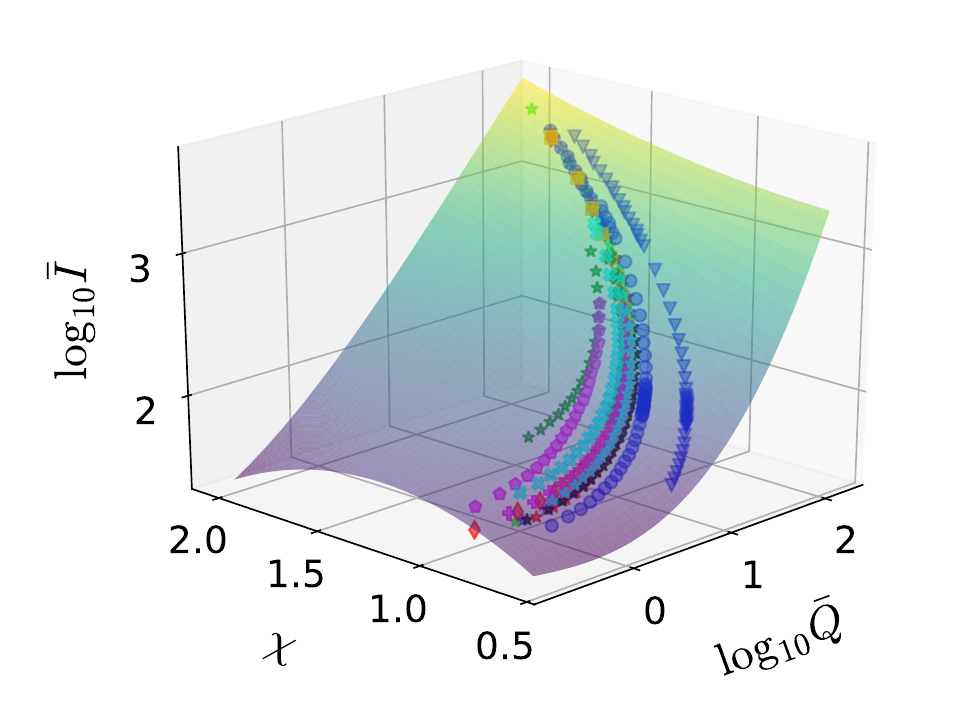}%
}

\subfloat{%
  \hspace*{1.0cm}\includegraphics[clip,width=0.94\columnwidth]{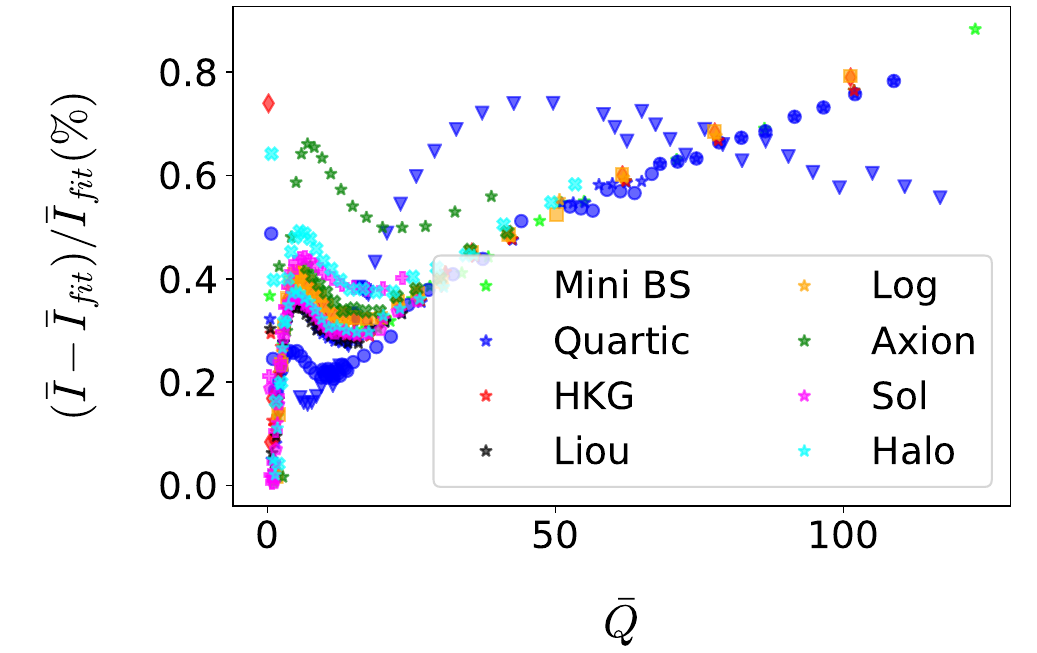}%
}
\caption{Universal $I-\chi-Q$ surface for spinning BSs fitting the data points obtained numerically (upper panel) and relative difference between data and fitted value, in percent (lower panel). Remarkably, the relation holds with an error of less than $1\%$. }
\label{super}
\end{figure*}

\begin{figure*}[]
\centering
\hspace*{0.5cm}\includegraphics[width=0.54\textwidth]{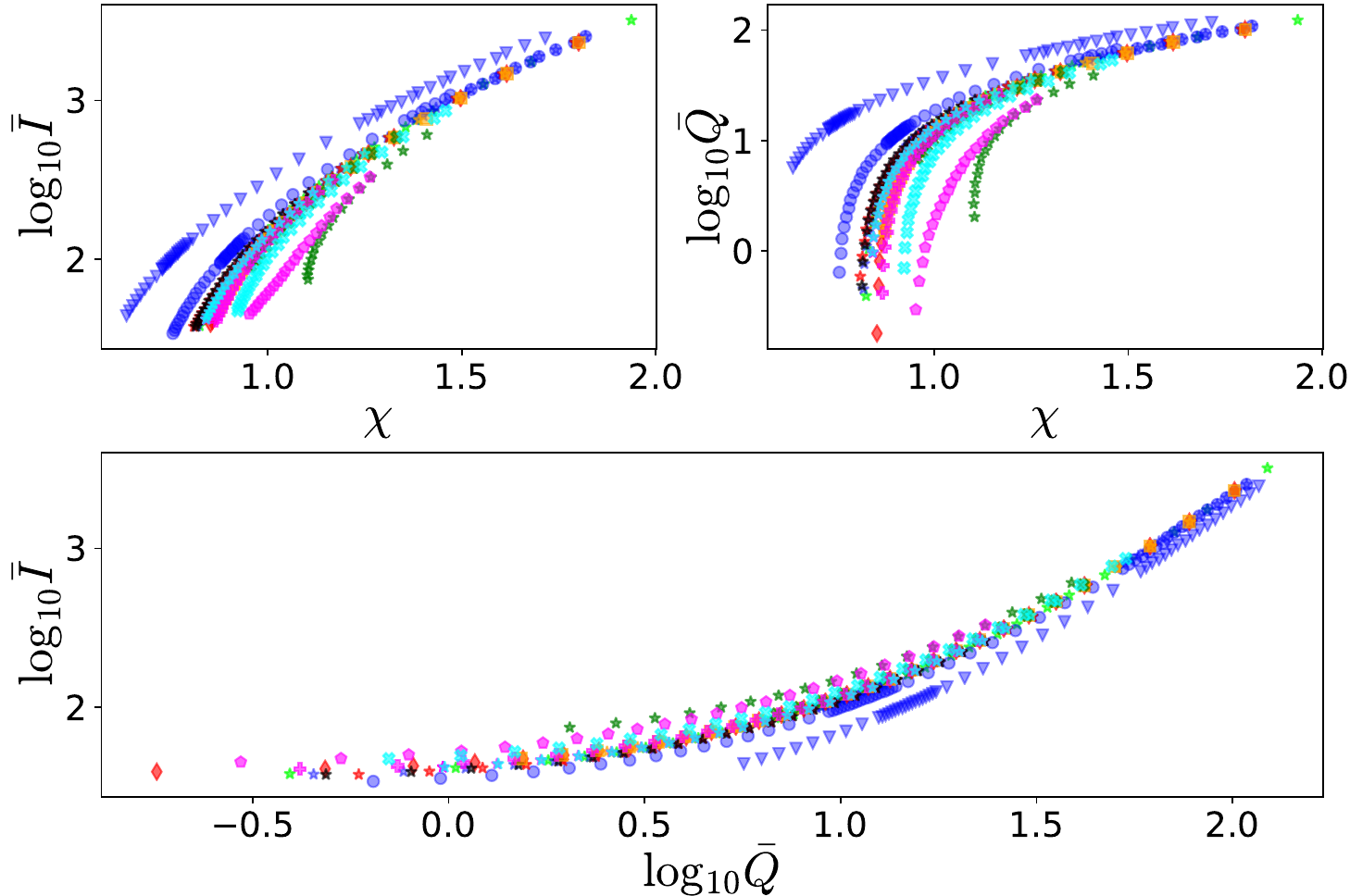}
\caption{We represent the same data as in \cref{super}, projected along each axis. We can appreciate from the 2D projections that we have to understand data in a 3D space parameter if we want to find the universal behavior, as happens for rotating NSs \cite{Pappas:2013naa}.  }
\label{projections}
\end{figure*}

For a more substantial improvement, we should take into account the spin frequency of the solutions, as in \cite{Pappas:2013naa} (see also \cite{Doneva:2017jop}). We consider our BS data in a 3D parameter space, where each point has coordinates $P(\bar{I},\bar{Q},\chi)$. If we represent our simulations in this 3D space, the moment of inertia can be seen as a surface function of the spin parameter and the quadrupole moment, i.e $I=F(Q,\chi)$. This surface can be fitted as,
\begin{equation}
    \begin{split}
       &\log_{10}{\bar{I}}=A_0+A_n^m\chi^m\left(\log_{10}\bar{Q}-B\right)^n,
    \end{split}
\end{equation}
\begin{table*}[h!]
	\centering
		\begin{tabular}{|c|c|c|}
			\hline
		 Coeffs & $A_0=1.3067$ & $B=  -0.7413$ \\ \hline
   $A_1^0= 0.0000$	& $A_1^1= 1.2793$  & $A_1^2= -0.7413$ \\ \hline
	$A_2^0=    -0.3078$& $A_2^1=-0.2426$ &  $A_2^2=  0.2678 $\\  \hline
	$A_3^0=  0.0796$ & $A_3^1=  -0.0045 $& $A_3^2= -0.0226$\\ \hline
		\end{tabular}
		\caption{Numerical values of the coefficients that fit the universal BSs $I\chi Q$ surface. }
		\label{Table.Constants}
\end{table*}
with $n={1,2,3}$, $m={0,1,2}$, and the fitting coefficients given in \cref{Table.Constants}.
The difference between the fitted surface and the real data is always less than $1\%$, see \cref{super,projections}. So the quadrupolar, angular and mass moments determine the moment of inertia with a very high precision in a model independent fashion.


An interesting comparison can be made between our fitted surface for rotating BSs and a similar result for rapidly rotating NSs \cite{Pappas:2013naa}. Indeed, the moments of inertia and quadrupole moments of spinning NSs were shown to follow a universal relation in the $(\bar{I},\bar{Q},\chi)$ parameter space as well, which yields a different $I-Q$ relation for fixed spin parameter. 
The main difference between rotating NSs and BSs is that in the first case we have enough freedom to fix the mass and $\chi$ independently, while in the second case, one of the two fixes the other. This means that for a concrete model, the inertia moment of NS solutions span a surface parametrized by $(\chi, \bar{Q})$ but for BSs they follow a single curve. Universality then comes from the fact that all data lie on the same surface, independently of the model or the parameter values. 

\begin{figure*}[]
\centering
\hspace*{-0.0cm}\includegraphics[width=0.50\textwidth]{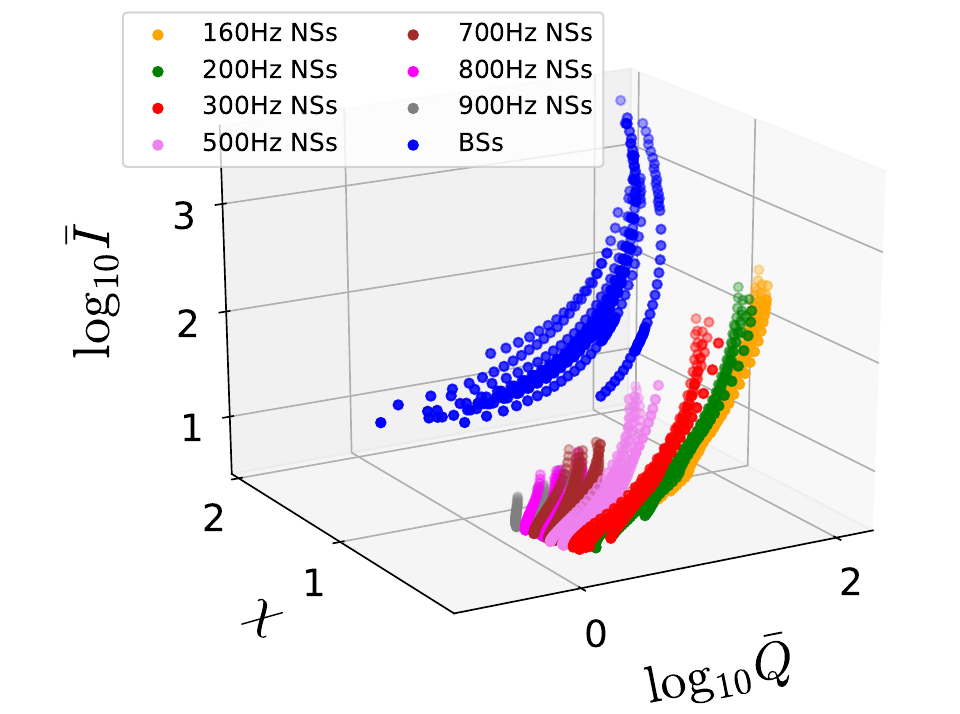}
\caption{Blue dots correspond to BS data. Colored points are NSs for different frequencies and several EOS, namely BCPM \cite{Sharma:2015bna}, AGHV \cite{Adam:2020yfv}, BPAL \cite{Zuo:1999vcl}, RNS-FPS \cite{Engvik:1994tj}, RNS-A \cite{Arnett:1977czg} and SLy \cite{Douchin:2001sv}. }
\label{NS}
\end{figure*}

In \cref{NS}, we compare our data set with several rapidly rotating NS solutions, for various EOS and angular velocities. Rapidly rotating NS data were obtained using the RNS package  \cite{stergioulas1992rotating}. The universal surface that the NS data form is clearly different from that generated by BSs. 

For similar quadrupole moments, the spin parameter space is larger for BSs than for NSs, as expected, and the moment of inertia of BSs is always higher. This can be understood from their different shapes, because a toroidal body typically has a higher moment of inertia than a spherical one of the same mass and equatorial radius.

Our results are interesting from a theoretical point of
view, as they confirm a particular universal behavior for
BSs.  They could also become important for observations, but this requires the simultaneous measurements of the moment of inertia, spin and quadrupole moments from GW observations alone, because standard BSs are not expected to produce observable signals in other (e.g., electromagnetic) channels of astrophysical observations.
Such simultaneous GW observations of $I$, $\chi$ and $Q$ are beyond current possibilities, but with sufficient progress both in the calculation of more detailed wave forms and in the precision of GW observations they may become possible in the future.

\section{Discussion and summary.}
In this work we present a universal relation between the reduced moment of inertia, spin parameter and reduced quadrupolar moment for rotating scalar BSs that is satisfied with a one percent accuracy for a great variety of bosonic potentials. 
For horizonless objects, such universal relations play a role similar to the no-hair theorem for black holes, because they allow us to determine the external gravitational field from a finite number of multipole moments with a high precision.
An interesting extension of our results would be to study universal relations for solutions with different values of the harmonic index and/or solutions in the limit in which a horizon has formed inside the rotating BS -- hairy Kerr black holes --  or the study of higher order multipoles like the spin octupole and mass hexadecapole.
As for their NS counterparts, we expect that these universal relations may become useful in the analysis of gravitational waveforms of future binary merger events, in the search of possible bosonic self-coupling terms for dark matter candidates and in the further understanding of the strong gravity regime of General Relativity.

We would like to end with several remarks. First of all,
 following \cite{Delgado:2020udb}, we use the perimetral radius $r_{99}$ which contains 99\% of the BS mass and the corresponding mass $M_{99}$ for our calculation of the reduced multipole moments. We checked, however, what happens if we use the full masses $M_{100}$, instead, and found that our results are not very sensitive to this change. Concretely, while the surface in the 
$I,\chi ,Q$ space slightly changes, its universality is maintained with the same, better than 1\%, precision. 

Secondly, we want to emphasize that
in \cite{Vaglio:2022flq} some multipolar moments for BSs for the quartic potential were studied in the  
 strong-coupling approximation of \cite{PhysRevD.55.6081}, and a comparison with our results would certainly be interesting, as we consider this potential, as well. The maximal value of the coupling constant we consider is, however, $\lambda /m^2 = 50$, whereas all results in \cite{Vaglio:2022flq} correspond to larger coupling constants. Further, the extension of our calculations to higher coupling constants requires some adaptions of our numerical methods. A comparative study will, therefore, be presented in a subsequent paper.

Thirdly, it is known \cite{Sanchis-Gual:2019ljs} that some of the BSs considered in this paper suffer from dynamical instabilities, such that after a sufficiently long time the rotating BS will either radiate away its angular momentum or collapse to a black hole. This instability can be avoided, e.g., by a sufficiently strong self-interaction of the boson field \cite{Sanchis-Gual:2021phr,Siemonsen:2020hcg,DiGiovanni:2020ror}. In other words, some of the rotating BSs considered here will be stable or long-lived, whereas others will have too short a life span to be of physical interest. The important point for us is that {\em all} these rotating BSs (i.e., for all models and all values of the coupling constants) span exactly the same universal surface in the $I-\chi -Q$ parameter space. That is to say, the universal relation is not affected by the (in-)stability of the corresponding BSs. 

Finally, all calculations in this paper were done for the harmonic index $n=1$, which defines a genuine surface in $I,\chi ,Q$ space. As already explained, for low $n > 1$, each $n$ defines its own universal surface in this space, which converges to a limiting surface in the limit $n>>1$. If $n$ is treated as a free (unknown) variable, therefore, we find a whole set of surfaces and  a whole set of values for $I$, say, for given $\chi ,Q$. In this situation, the values of $\chi ,Q$ are no longer sufficient to predict the value of $I$, but the resulting region in $I,\chi ,Q$ space can still be clearly distinguished from the values provided by other types of compact stars, like NSs or BHs.

\begin{acknowledgements}
The authors thank C. Naya for helpful discussions. JCM thanks E.Radu and J.Delgado for their crucial help with the FIDISOL/CADSOL package. 
Further, the authors acknowledge financial support from the Ministry of Education, Culture, and Sports, Spain (Grant No. PID2020-119632GB-I00), the Xunta de Galicia (Grant No. INCITE09.296.035PR and Centro singular de investigación de Galicia accreditation 2019-2022), the Spanish Consolider-Ingenio 2010 Programme CPAN (CSD2007-00042), and the European Union ERDF.
AW is supported by the Polish National Science Centre,
grant NCN 2020/39/B/ST2/01553.
AGMC is grateful to the Spanish Ministry of Science, Innovation and Universities, and the European Social Fund for the funding of his predoctoral research activity (\emph{Ayuda para contratos predoctorales para la formaci\'on de doctores} 2019). MHG and JCM thank the Xunta de Galicia (Consellería de Cultura, Educación y Universidad) for the funding of their predoctoral activity through \emph{Programa de ayudas a la etapa predoctoral} 2021. JCM thanks the IGNITE program of IGFAE for financial support.
\end{acknowledgements}

\appendix
\section{ Numerical parameters used.}

Here we give the different numerical sets of values for the parameters that we have used for our simulations. The numerical values are given in rescaled units.
\begin{equation}
V_{\rm Quartic}=\phi^2+\frac{\lambda}{2}\phi^4
\begin{cases}
      \lambda= 1\\
      \lambda=10\\
      \lambda=50
\end{cases}
 \end{equation}

\begin{equation}
V_{\rm Halo}=\phi^2-\alpha\phi^4
\begin{cases}
      \alpha= 1,\\
      \alpha=12.\\
\end{cases}
 \end{equation}

\begin{equation}
V_{\rm HKG}=\phi^2-\alpha\phi^4+\beta\phi^6
\begin{cases}
      \alpha=80, &\beta=0.01\\
      \alpha=2, &\beta=1.8\\
\end{cases}
 \end{equation}

\begin{equation}
V_{\rm Sol}=\phi^2\left(1-\left(\frac{\phi^2}{\phi_0^2}\right)\right)^2
\begin{cases}
      \phi_0=1.5, \\
      \phi_0=0.7,\\
      \phi_0=0.3.
\end{cases}
 \end{equation}

\begin{equation}
\begin{split}
V_{\rm Axion}&=\frac{2f^2}{B}\left(1-\sqrt{1-4B\sin^2(\frac{\phi}{2f})}\right)\\&
\begin{cases}
      f=0.1, &B=0.22.\\
      f=0.05, &B=0.22.
\end{cases}
\end{split}
 \end{equation}

\begin{equation}
V_{\rm Log}=f^2 \ln\left(\phi^2/f^2+1\right)
\begin{cases}
      f=0.7, \\
      f=0.5.\\
\end{cases}
 \end{equation}

\begin{equation}
V_{\rm Liouville}=f^2 \left(e^\frac{\phi^2}{f^2}-1\right)
\begin{cases}
      f=0.8. \\
\end{cases}
 \end{equation}

\section{Precision and numerical stability.}

\begin{figure*}
    \centering
    \begin{subfigure}[t]{0.25\textwidth}
        \centering
        \includegraphics[width=\linewidth]{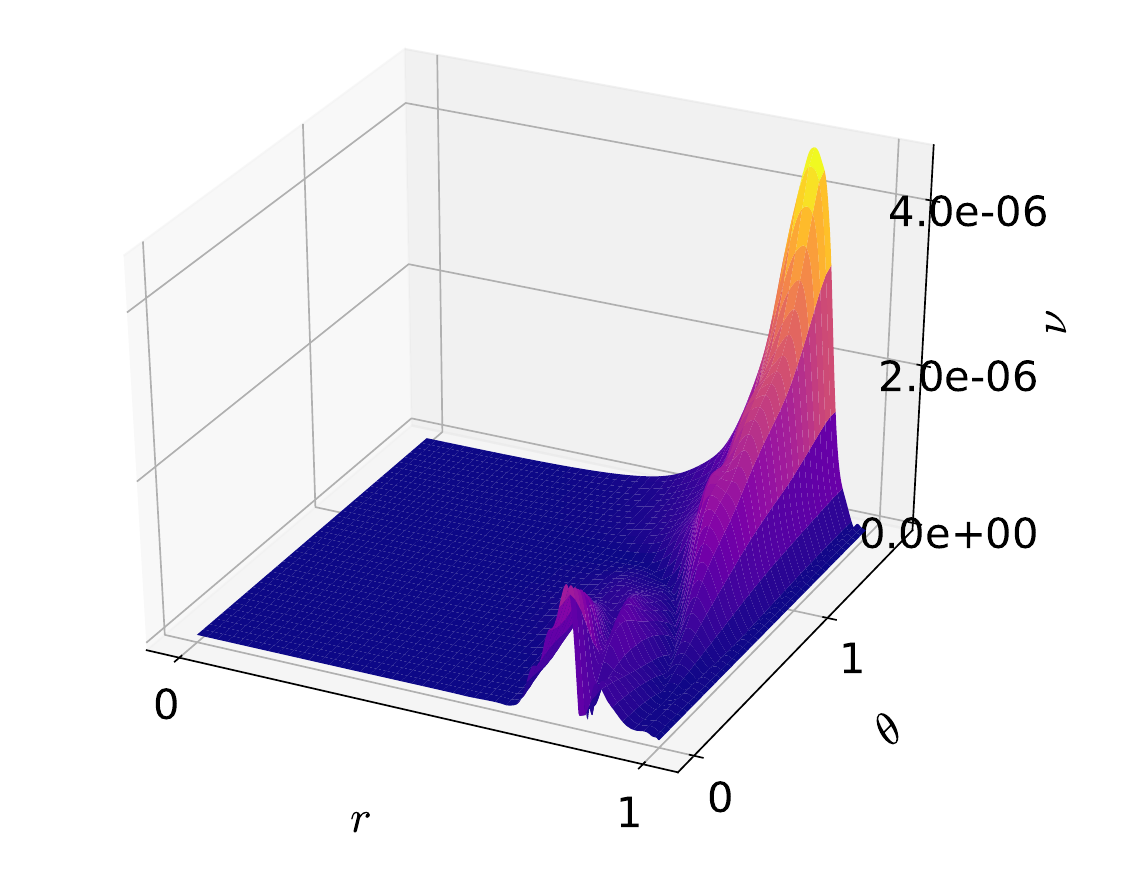} 
        \caption{Residuals $\nu$} \label{fig:timing1}
    \end{subfigure}
    \hfill
    \begin{subfigure}[t]{0.25\textwidth}
        \centering
        \includegraphics[width=\linewidth]{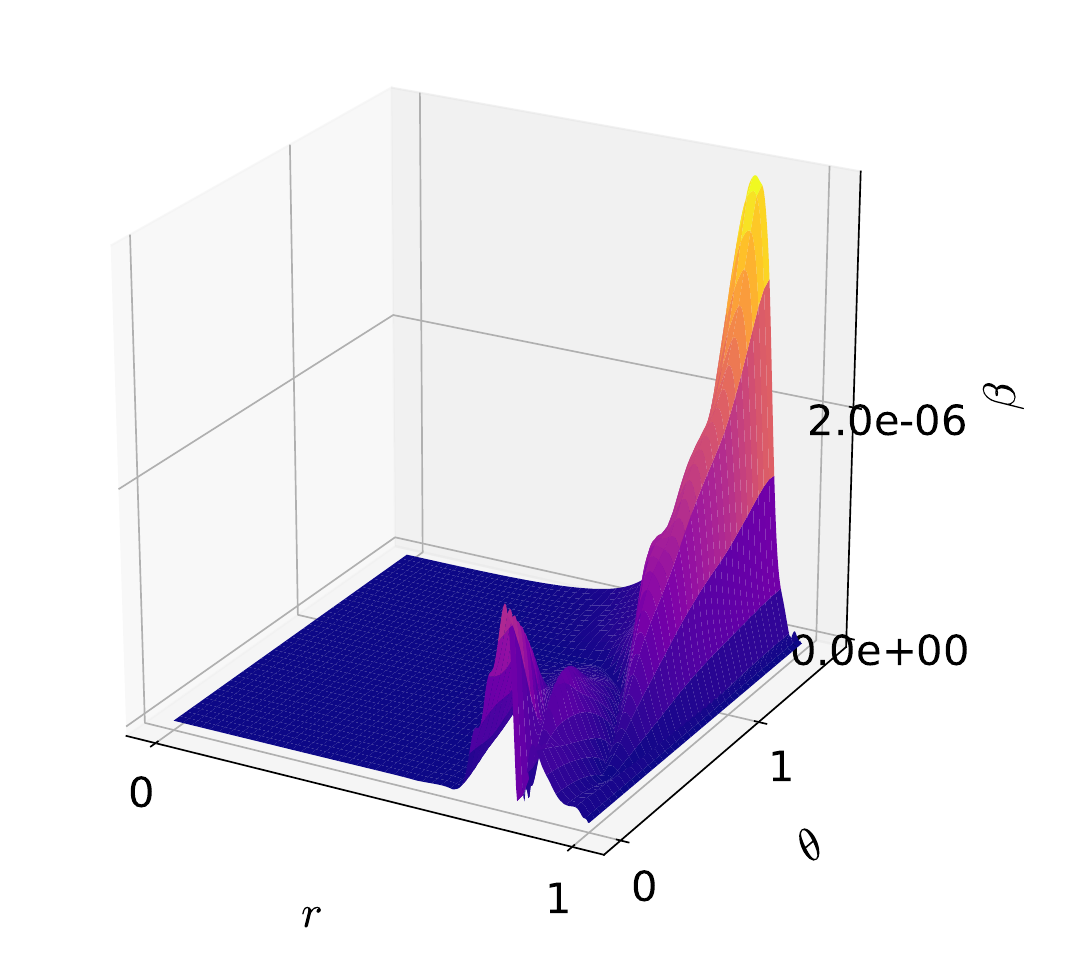} 
        \caption{Residuals $\alpha$} \label{fig:timing2}
    \end{subfigure}
    \centering
    \begin{subfigure}[t]{0.25\textwidth}
        \centering
        \includegraphics[width=\linewidth]{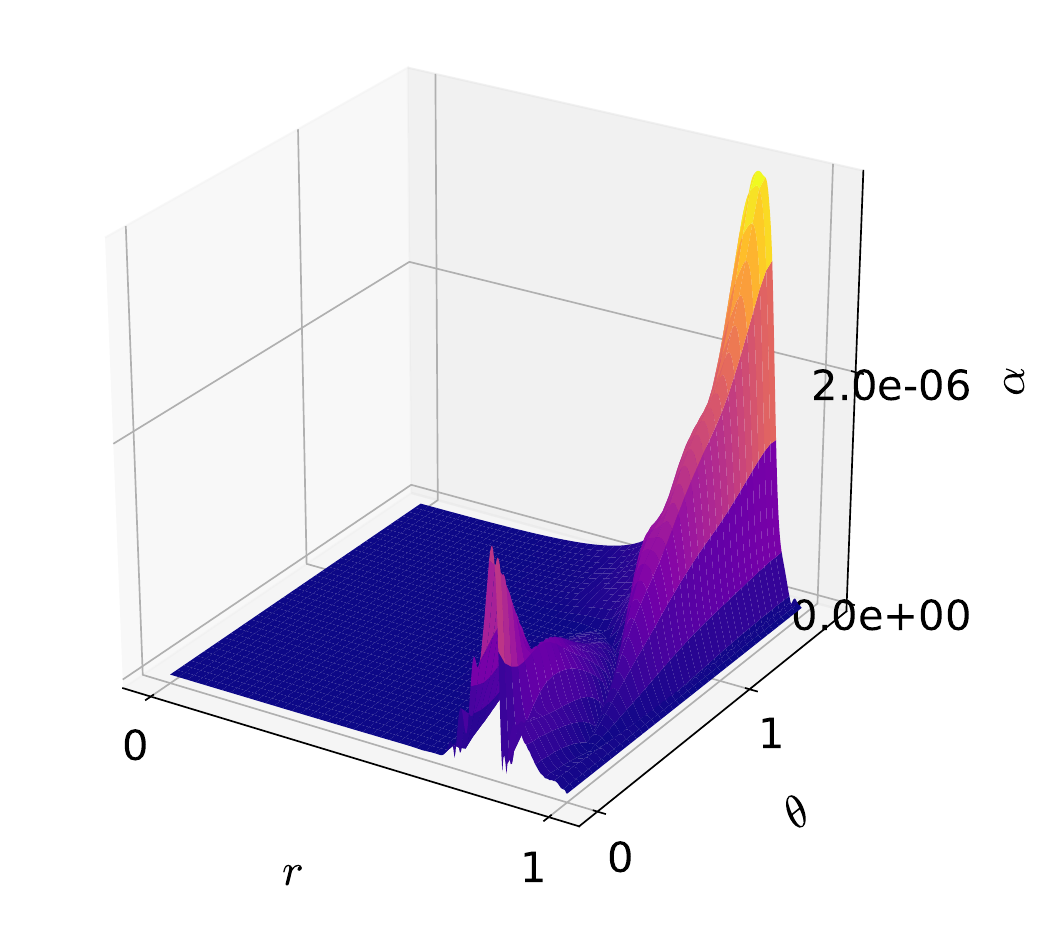} 
        \caption{Residuals $\beta$} \label{fig:timing1}
    \end{subfigure}
    \hfill
    \begin{subfigure}[t]{0.25\textwidth}
        \centering
        \includegraphics[width=\linewidth]{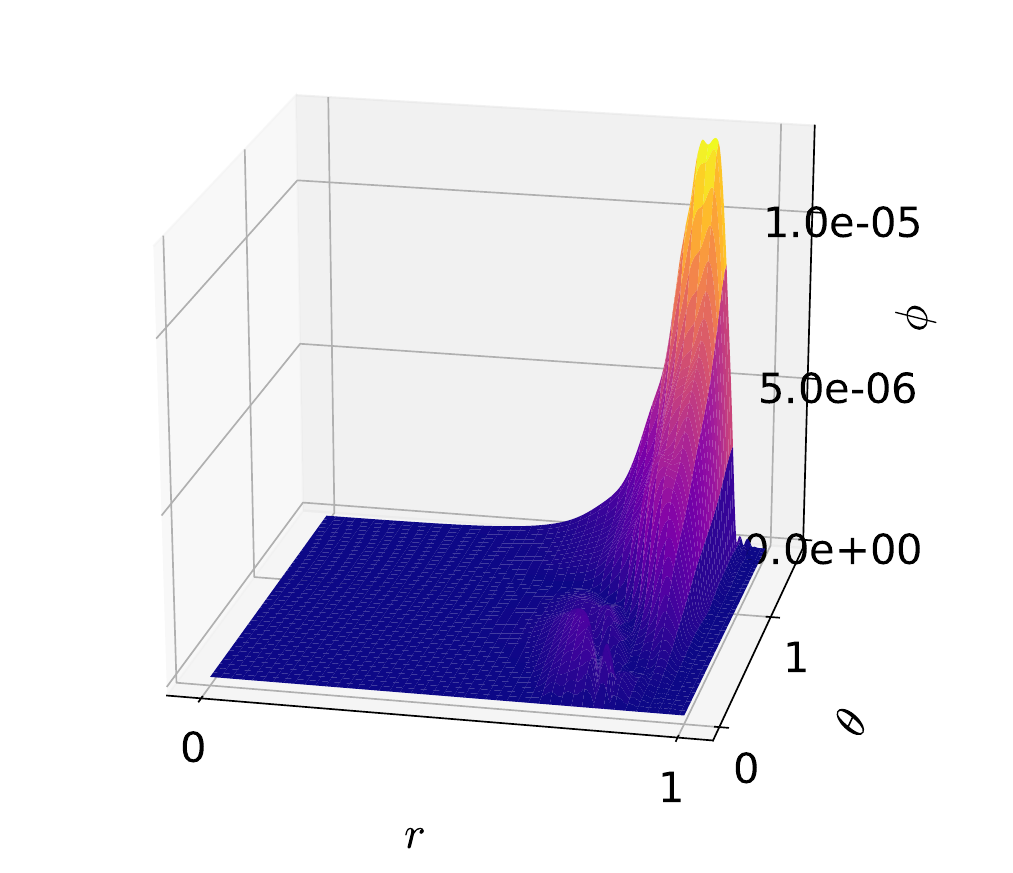} 
        \caption{Residuals $\phi$} \label{fig:timing2}
    \end{subfigure}
    \begin{subfigure}[t]{0.25\textwidth}
    \centering
        \includegraphics[width=\linewidth]{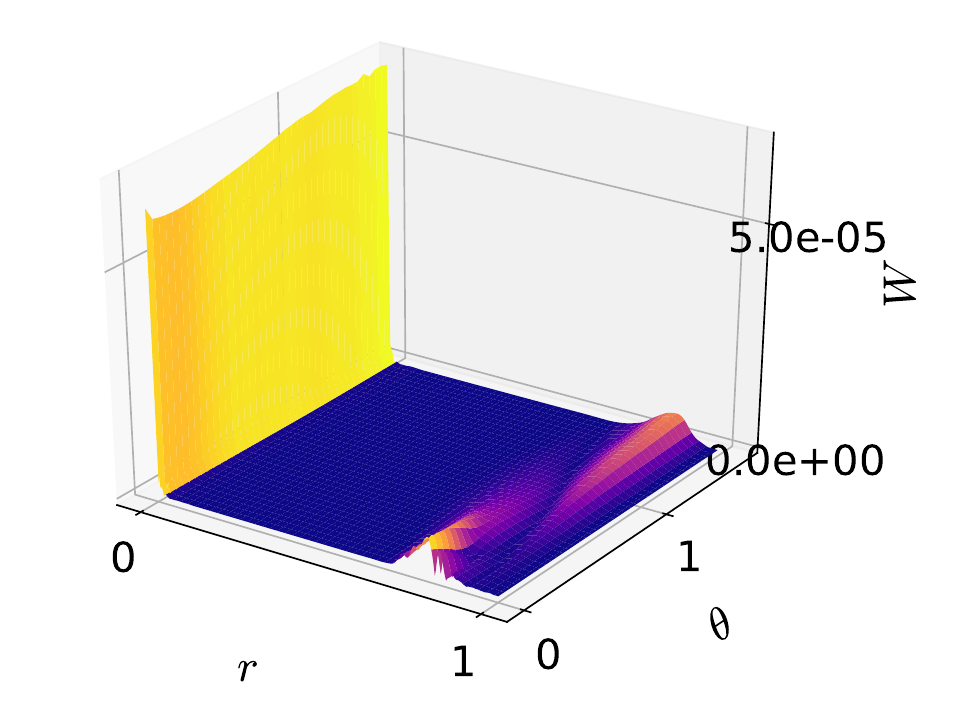} 
        \caption{Residuals $W$} \label{fig:timing3}
    \end{subfigure}
    \caption{}
    \label{Figura}
\end{figure*}

The precision of our numerical calculations for a fixed, given grid is controlled by the Newton residuals of the different fields. For our standard $401 \times 40$ grid, we show these residuals in 
\cref{Figura} for the particular case of the rotating mini-boson star (with the mini-boson star potential). 
It can be appreciated that for the metric potentials $\nu,\beta,\alpha$ and the field $\phi$ the Newton Residuals behave quite similarly. The maximum values for the residuals occur for $x=\Bar{r}=1$ and  $y=\theta=\pi/2$. There is a second peak at the same radius and $\theta=0$ but with a lower value. 
In any case, all residuals are much below the required tolerance. The $W$ residuals, on the other hand, behave differently. At $\Bar{r}=0$ and for $\theta$ taking all the range $[0,\pi/2]$, the values are much higher than for the rest of the grid. 
This is probably related to the fact that the boundary value $W(r=0, \theta )=0$ is imposed exactly in our numerical integration, whereas the other fields are allowed to take their boundary values numerically.
But even the higher residuals of $W(r, \theta)$  are always smaller than the required tolerance.

\begin{figure*}
    \centering
    \begin{subfigure}[t]{0.25\textwidth}
        \centering
        \includegraphics[width=\linewidth]{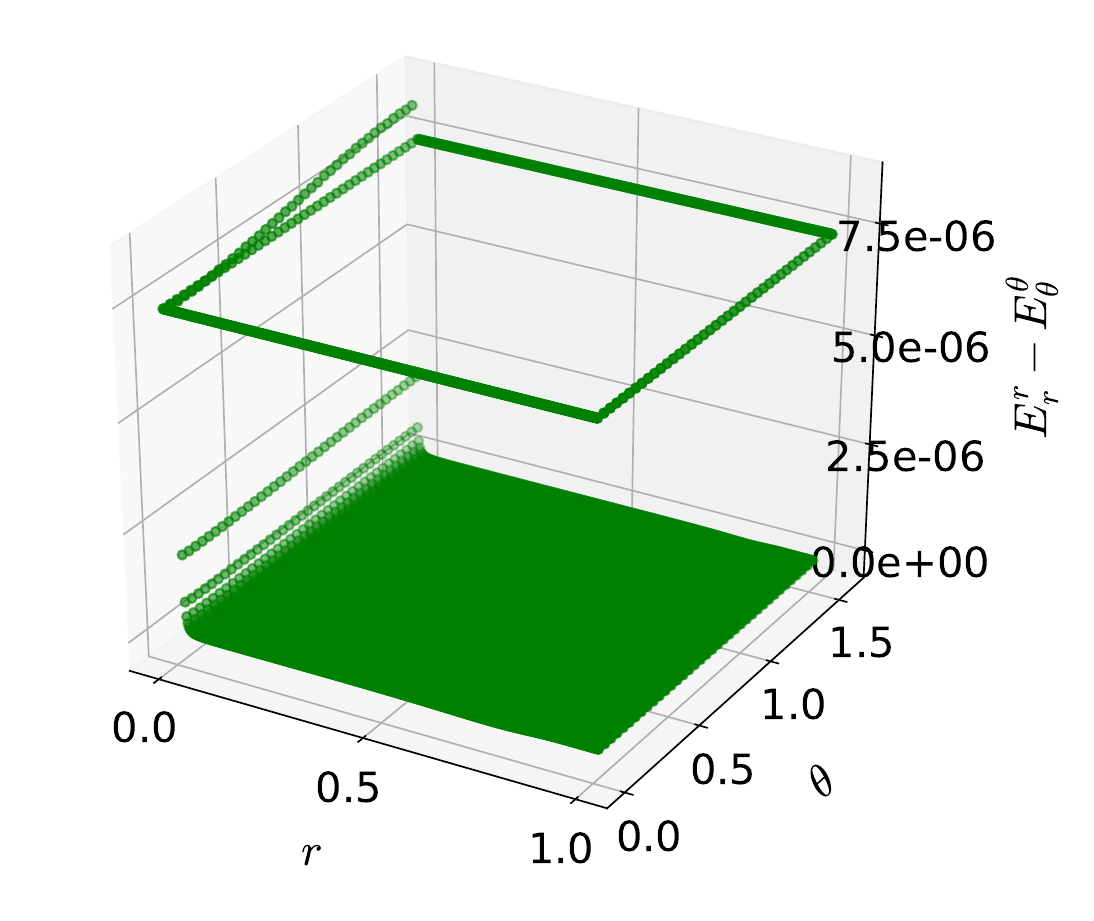} 
        \caption{Constraint equation $E^r_r-E^{\theta}_{\theta}=0.$} \label{fig:timing1}
    \end{subfigure}
    \hfill
    \begin{subfigure}[t]{0.25\textwidth}
        \centering
        \includegraphics[width=\linewidth]{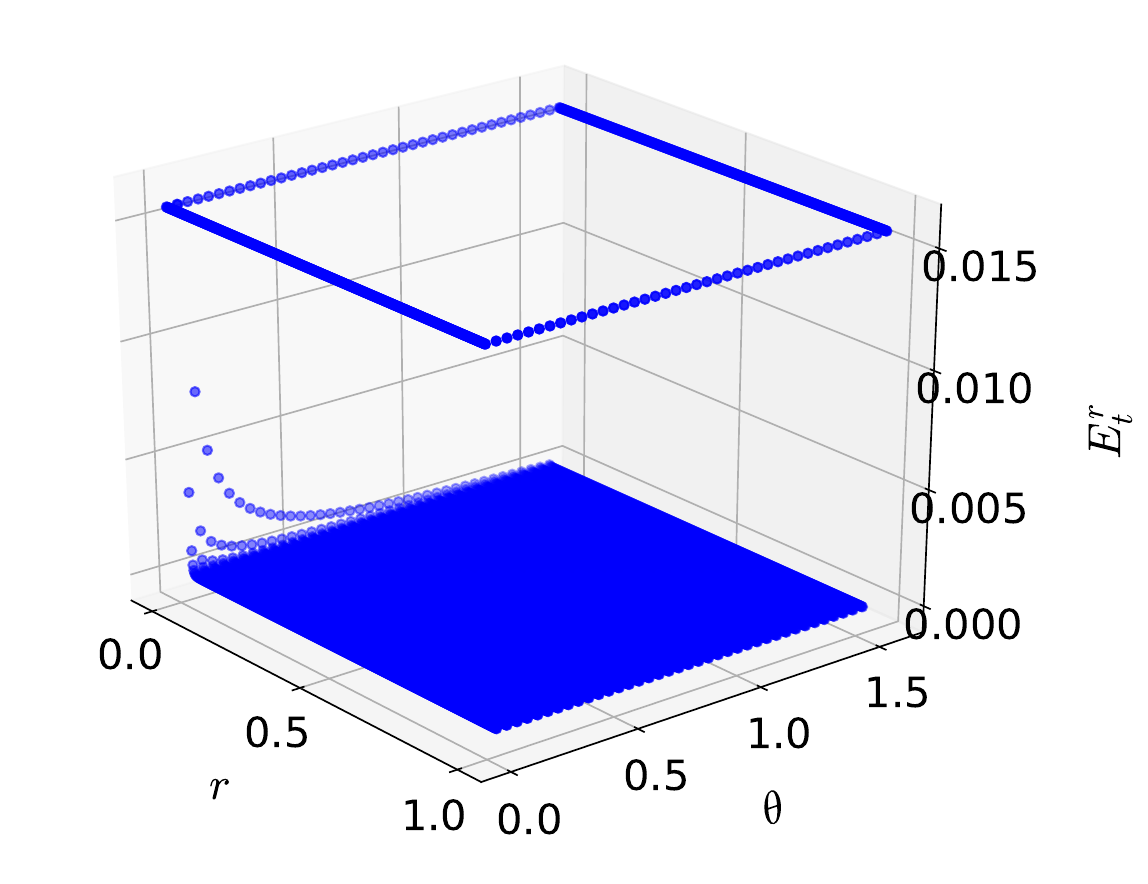} 
        \caption{Constraint equation $E^{\theta}_r=0.$} \label{fig:timing2}
    \end{subfigure}
    \caption{}
\label{figura2}    
\end{figure*}

To check the physical reliability of our results we use certain physical constraints which the system must obey. Indeed, in addition to the (second order) equations \cref{EKG-system}, the following constraint equations must hold,
\begin{equation}
\begin{split}
  &\text{Con1}\hspace{0.8cm}  E^r_r-E^{\theta}_{\theta}=0,\\
  &
  \text{Con2}\hspace{0.8cm}    E^r_t=0.
\end{split}
\label{constr}
\end{equation}
We plot the above equations for each point of the grid in \cref{figura2}.
We can observe that in the interior of the grid both constraints hold with a precision of
better than $ 10^{-6}$. But when we reach the border of the grid, the deviations increase several orders of magnitude, in particular for Con2.  We know, however, that this is due to border effects, where the derivatives of some of the metric potentials are increased, therefore we can still be confident with our calculations. There are many further physical checks, like the Virial integrals, or the differences of ADM vs Komar quantities, but these go beyond the scope of the present letter. 

Finally, to check the numerical stability of our calculations, we repeated them for different grid sizes for the particular case of the mini-boson star potential. Concretely, we chose the following grids,
 \begin{table}[h!]
\begin{tabular}{|l|l|l|l|l|l|l|l|l|}
\hline
 $N_x$&135  &135  & 201 &201  &268  &310  &401    \\ \hline
 $N_y$&35  & 40 & 30 &40  &40  &40  &40  \\ \hline
\end{tabular}
\caption{\small Different grids, being $N_x$ the number of radial points and $N_y$ the number of angular points.}
		\label{grids}
\end{table}

\begin{figure}
    \centering
    \begin{subfigure}[t]{0.25\textwidth}
        \centering
        \includegraphics[width=\linewidth]{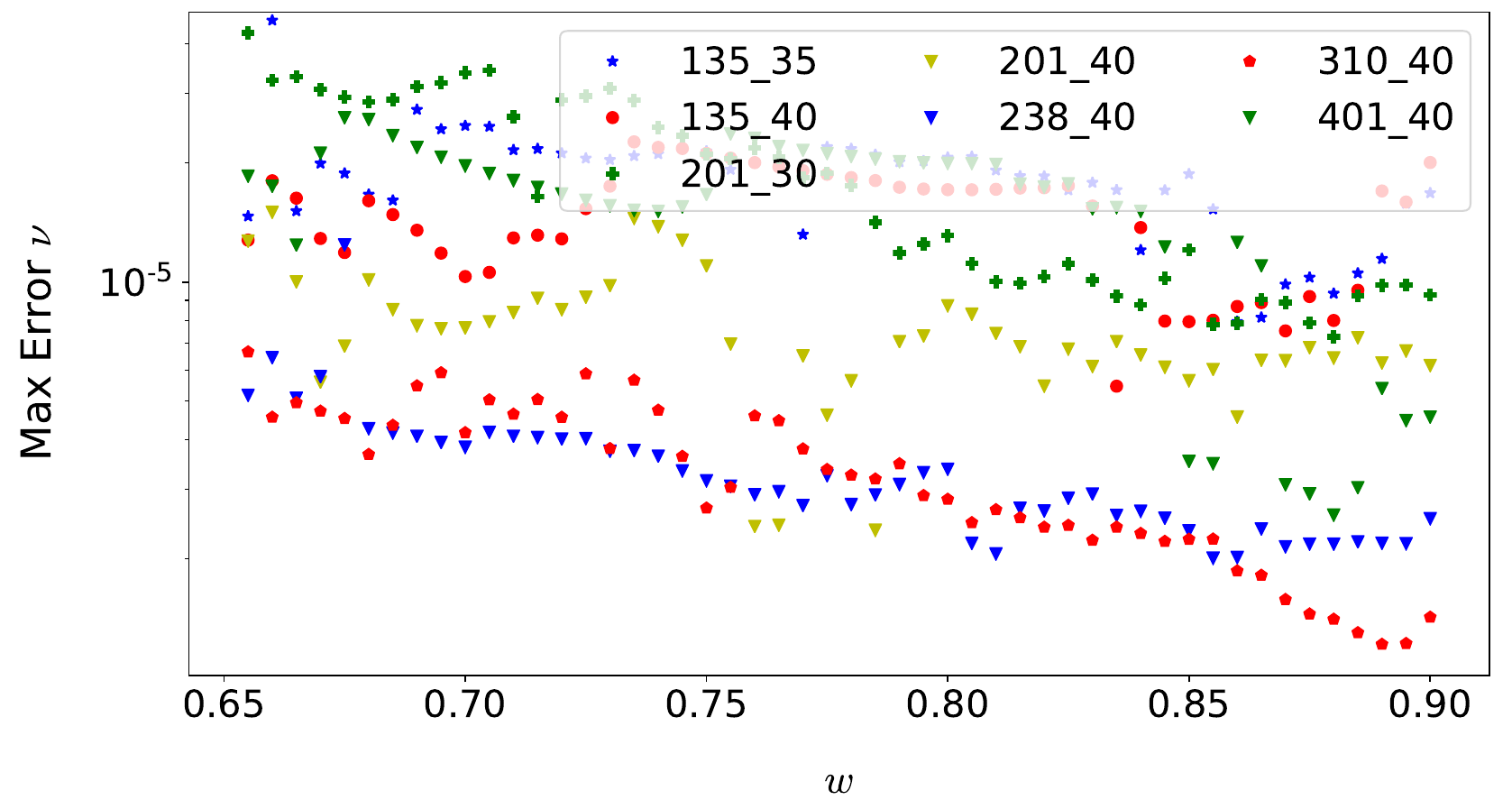} 
        \caption{Max residuals $\nu$} \label{fig:timing1}
    \end{subfigure}
    \hfill
    \begin{subfigure}[t]{0.25\textwidth}
        \centering
        \includegraphics[width=\linewidth]{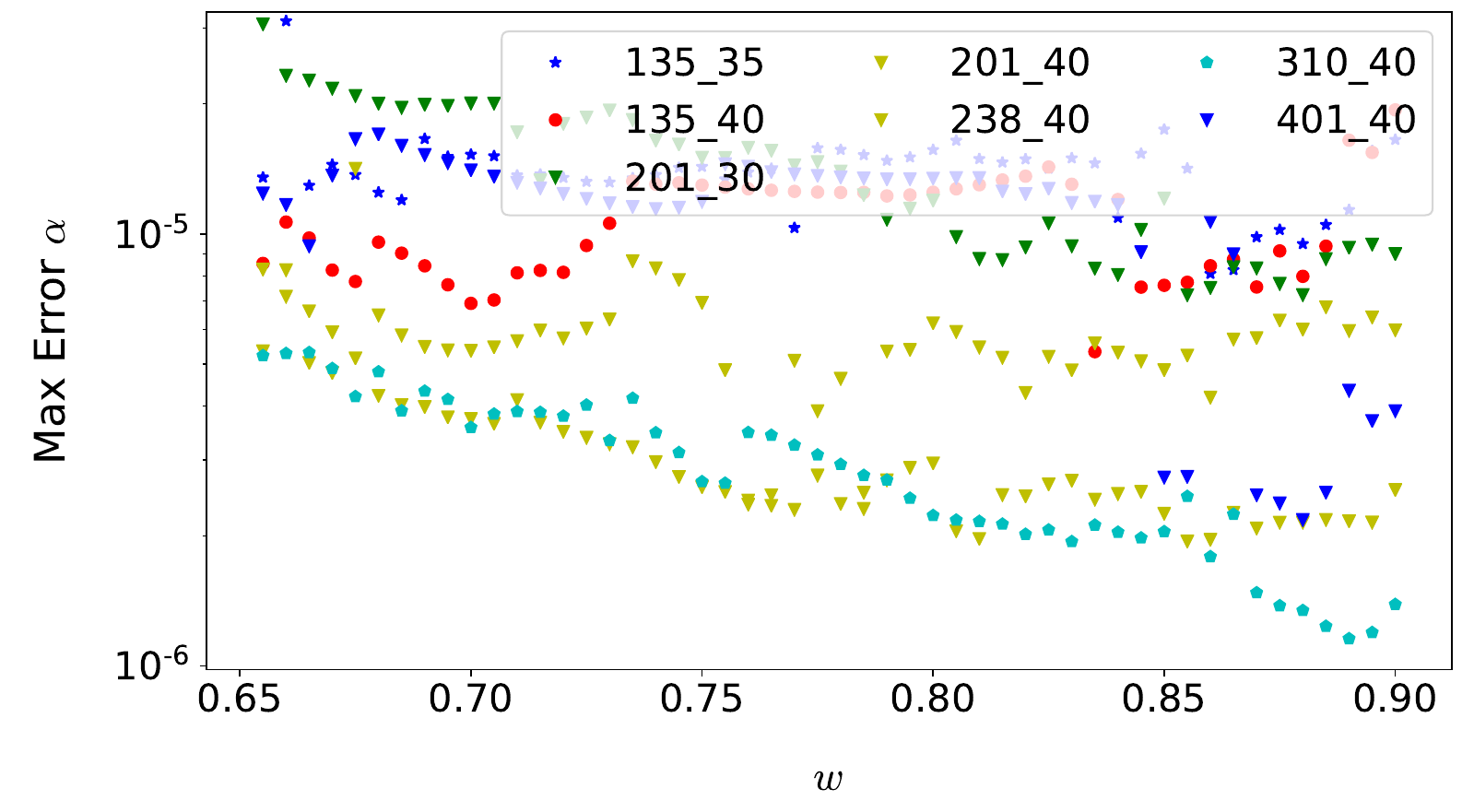} 
        \caption{Max residuals $\alpha$} \label{fig:timing2}
    \end{subfigure}
    \centering
    \begin{subfigure}[t]{0.25\textwidth}
        \centering
        \includegraphics[width=\linewidth]{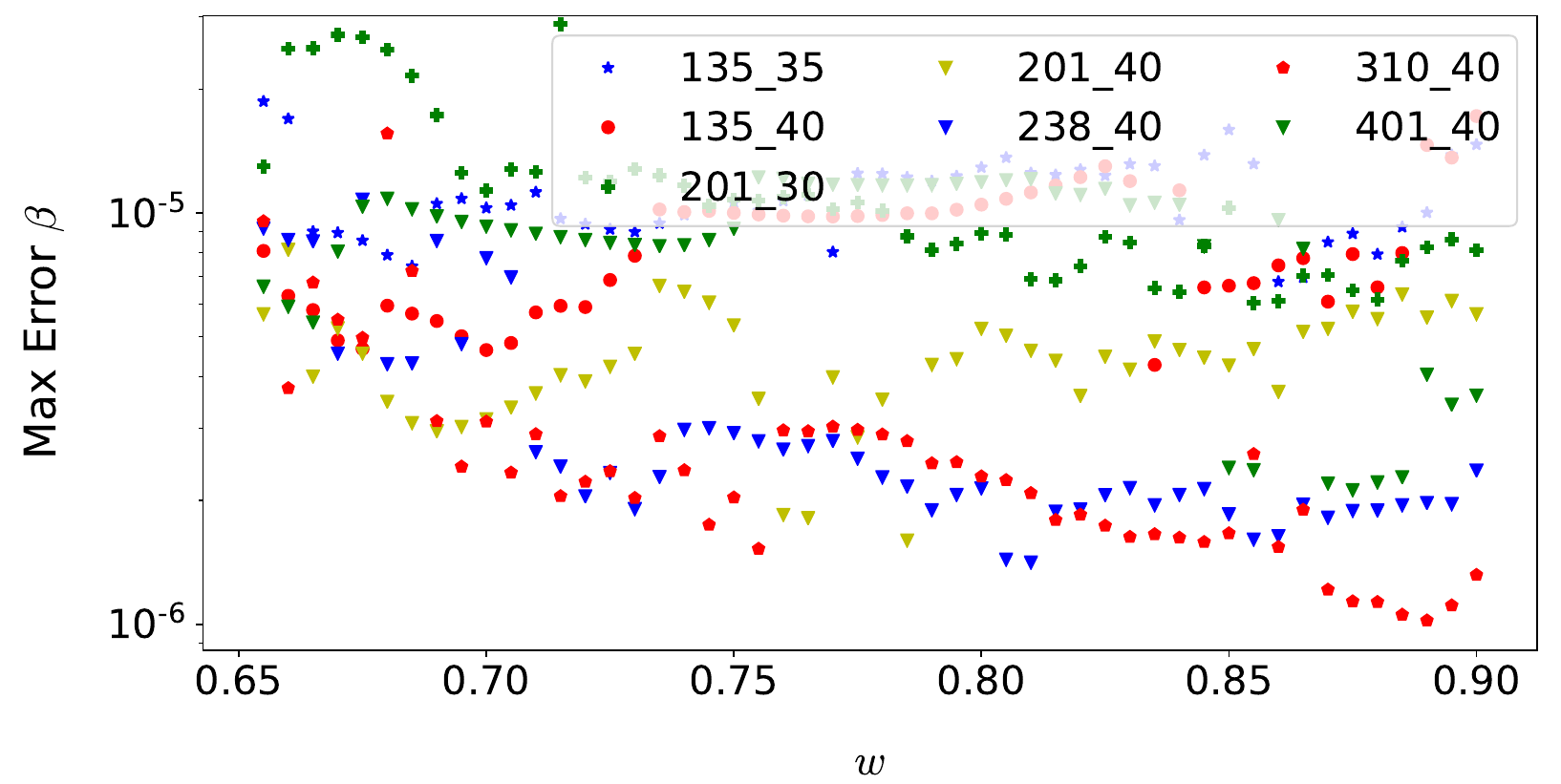} 
        \caption{Max residuals $\beta$} \label{fig:timing1}
    \end{subfigure}
    \hfill
    \begin{subfigure}[t]{0.25\textwidth}
        \centering
        \includegraphics[width=\linewidth]{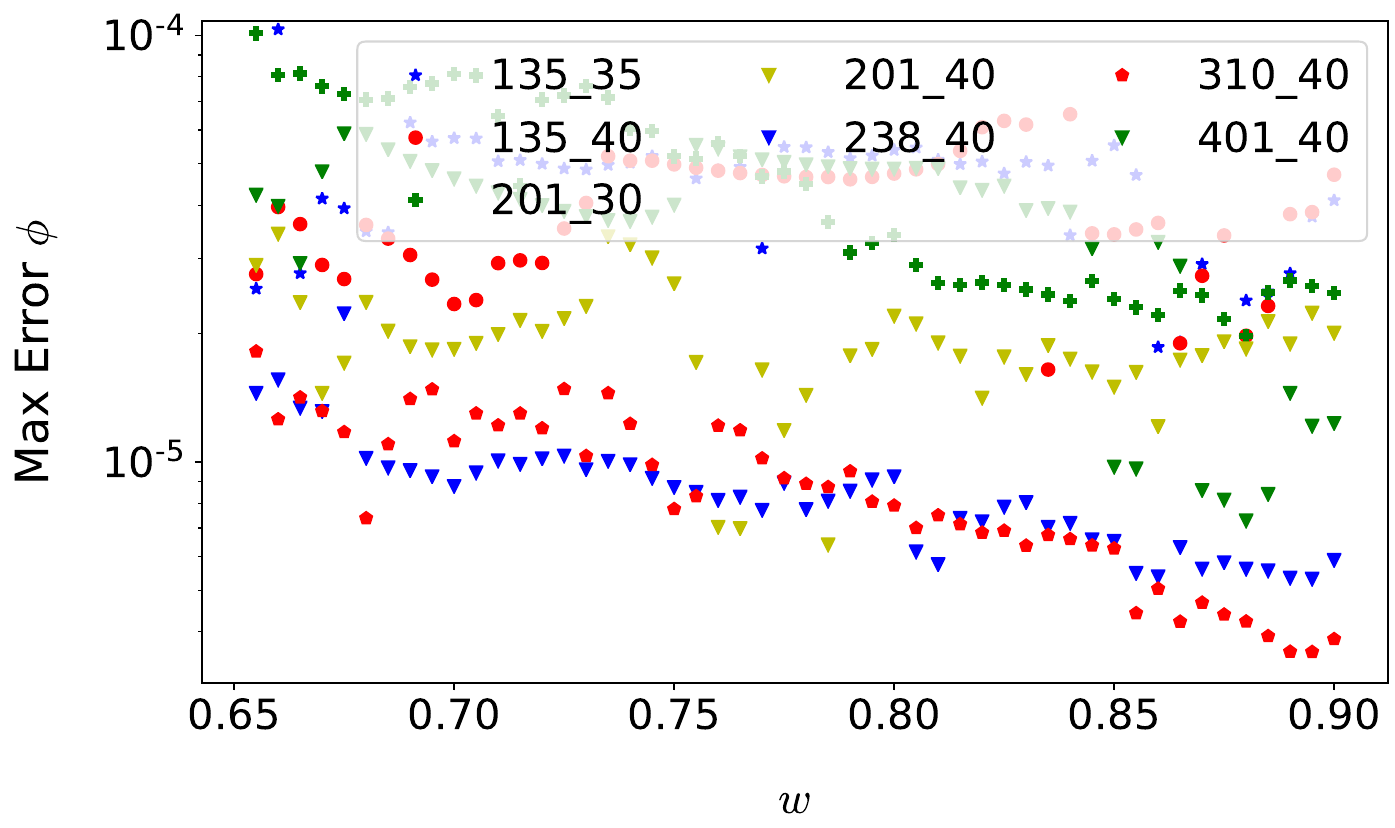} 
        \caption{Max residuals $\phi$} \label{fig:timing2}
    \end{subfigure}
    \begin{subfigure}[t]{0.25\textwidth}
    \centering
        \includegraphics[width=\linewidth]{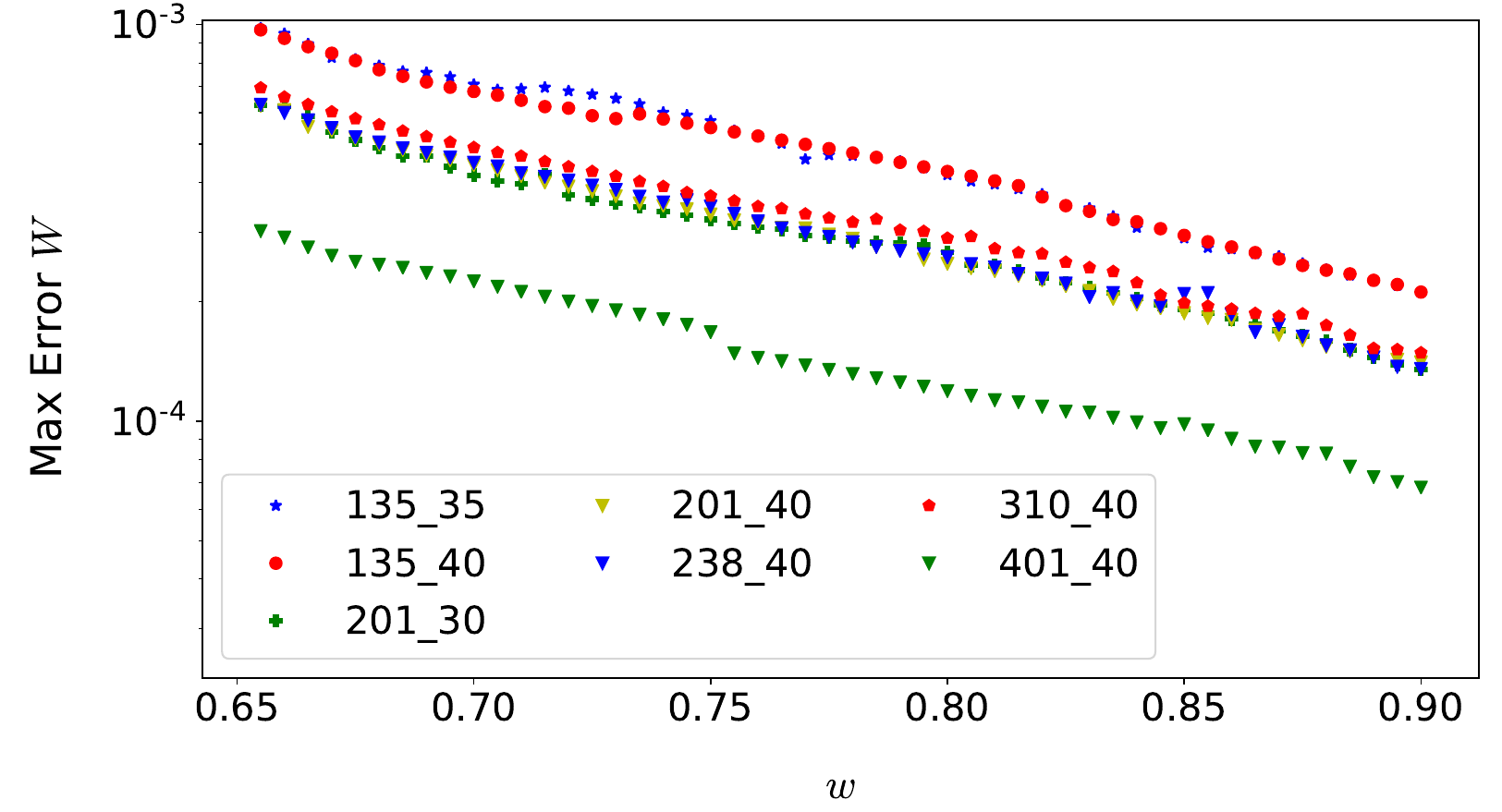} 
        \caption{Max residuals $W$} \label{fig:timing3}
    \end{subfigure}
    \caption{}
    \label{Figura3}
\end{figure}

 As is common in finite difference methods, the grid size can be changed
 only within a certain range. For too large grids, the memory allocation for the additional points is no longer possible, and the treatment of such large grids would require a significant rewriting of the code.
 We did the simulations for several field frequencies, from close to the minimum frequency in the main branch, $w= 0.655$, to the usual Newtonian region $w=0.9$. We did so because, as we can see in \cref{Figura3}, for different regions in $w$ the residuals will change, as happens for other quantities like the computation time. 
 
In \cref{Figura3} we show the maximum value of the residuals for each field. 
For the first four fields, all the grids returned maximal errors which are smaller than $10^{-4}$. Besides, if we considered only those quantities, we would prefer the grids $N_x=268, N_y=40$ or $N_x=310, N_y=40$, due to the lesser errors and faster computation times. But the metric potential that determines our choice of the grid is $W$, as it is the most sensitive. As already mentioned, due to the boundary conditions there exists a steep increase in the errors when $\Bar{r}=0$. For $W$, the grid that maintains the best (smallest) residuals is $N_x=401, N_y=40$. 
If we ignored or eliminated the rather large residuals of $W$ at the boundary $\Bar{r}=0$, then a better option for the grid would probably be $N_x=268, N_y=40$, because the simulation time would be shorter and the residuals in general would be lower.

\begin{figure}[h!]
\centering
\includegraphics[width=0.4\textwidth]{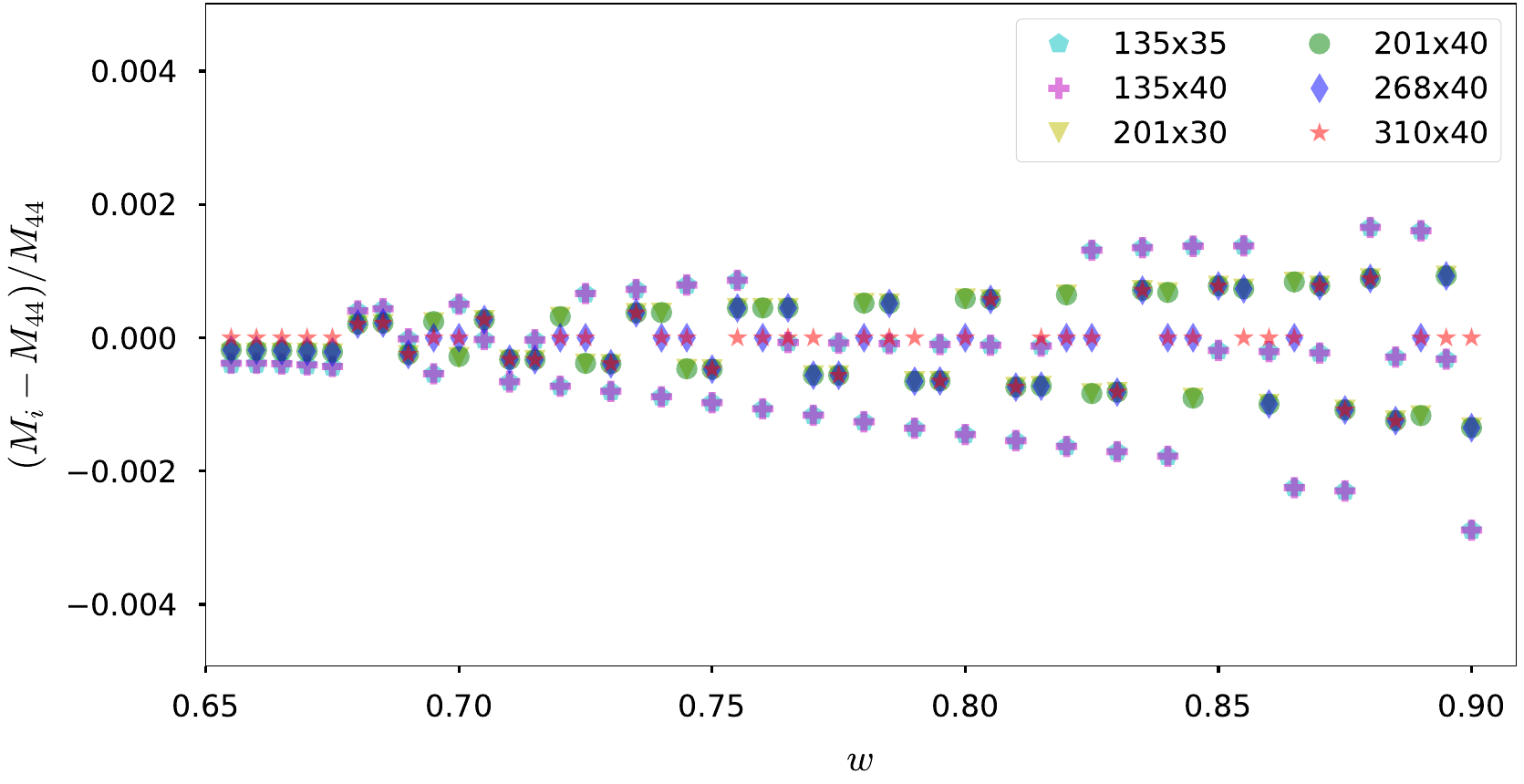}
\caption{Relative mass differences for different grids as a function of the frequency. Here $M_{44}$ is the mass of the corresponding rotating mini-boson star for our standard grid $401\times 40$, whereas $M_i$ are the masses for the other grids.  
}
\label{mass-dif-grid}
\end{figure}

Finally, in \cref{mass-dif-grid} we plot the rotating mini-boson star masses as a function of the frequency, for the different grids we used. In all cases, 
the relative differences to our standard grid $401\times 40$ are always less than $3\cdot10^{-3}$. The results for other observables are similar. These differences are slightly bigger than the errors for the fields ($\le 10^{-3}$), because the numerical integration needed for the calculation of the observables from the fields and their derivatives introduces some additional small errors.

\end{document}